\pdfoutput=1
\documentclass[useAMS,usenatbib,usegraphicx]{mn2e}
\usepackage{multirow}
\usepackage{url}
\usepackage{graphicx}
\def\PTP{{\it Prog.\ Theor.\ Phys.}}
\def\PNAO{{\it Publ.\ National Astron.\ Obs.\ Japan}}

\def\RS{{$R_{\it{s}}$}}
\def\DEG{^{\circ}}
\def\SGR{{SgrA$^{*}$}}
\def\PC{\%}
\def\UVC{\textit{u-v}}
\newcommand{\etal}{et al.}
\newcommand{\muas}{\mbox{$\mu$as}}

\title[Oscillation in \SGR]
{Oscillation phenomena in the disk around the massive black hole Sagittarius A$^{*}$}
\author[M. Miyoshi et al.]
{M.~Miyoshi,$^1$\thanks{E-mail address: \texttt{makoto.miyoshi@nao.ac.jp}},
Zhi-Qiang Shen$^{2}$,
T. Oyama$^{1}$,
R. Takahashi$^{3}$,
Y. Kato$^{4}$ \\
$^1$ National Astronomical Observatory of Japan, Mitaka, Tokyo, 181-8588, Japan \\
$^2$ Shanghai Astronomical Observatory, 80 Nandan Road, Shanghai, 200030, China \\ 
$^3$ The Institute of Physical and Chemical Research (RIKEN), 2-1 Hirosawa, Wako, Saitama, 351-0198, Japan \\
$^4$ Institute of Space and Astronautical Science (ISAS),
Japan Aerospace Exploration Agency (JAXA), \\
3-1-1 Yoshinodai, Sagamihara, Kanagawa, 229-8510, Japan \\
}
\date{Accepted --. Received 2010 January 12; in original form 2009 June 30.}
\pagerange{\pageref{firstpage}--\pageref{lastpage}} \pubyear{20**}

\begin{document}
\label{firstpage}
\maketitle 

\begin{abstract}
 We report the detection of radio QPOs with structure changes using the Very Long Baseline Array (VLBA) at $43$ GHz.
 We found conspicuous patterned changes of the structure with 
 $P = 16.8 \pm 1.4, 22.2 \pm 1.4, 31.2 \pm 1.5, 56.4 \pm 6$ min
very roughly in a $3:4:6:10$ ratio. The first two periods show a rotating one-arm structure, while the $P = 31.4$ min shows a rotating 3-arm structure, as if viewed edge-on. At the central $50\muas$ the $P = 56.4$ min period shows a double amplitude variation of those in its surroundings.
 Spatial distributions of the oscillation periods suggest that 
 the disk of \SGR ~is roughly edge-on, rotating around an axis with $PA = -10\DEG$.
 Presumably, the observed VLBI images of \SGR ~at 43 GHz retain several features
of the black hole accretion disk of \SGR ~in spite of being obscured and broadened 
by scattering of surrounding plasma.
\end{abstract}

\begin{keywords}
QPO: general,
QPO: individuals: \SGR
\end{keywords}

\section{INTRODUCTION}
 The existence of black holes has been definitely established (Miyoshi et al. 1995; and Harms et al. 1994) while zooming-in the relativistic region is still in difficulty~(Tanaka et al. 1995) though promising in near future (Falcke et al. 2000, Miyoshi et al. 2004, 2007, Doeleman et al. 2008, Fabian 2009).
 Sagittarius A$^{*}$ (\SGR), the most convincing massive black hole (Ghez et al. 2001; Sch\"odel et al. 2002, and Shen et al. 2005) at the Galactic center, shows short time flares with quasi-periodic oscillations (QPO) with $P = 17, 22, \&~33$ min in near-infrared and X-ray regions originated from near the central black hole (Baganoff et al. 2000; Genzel et al. 2003; Aschenbach et al. 04; Eckart et al. 2006; Belanger et al. 2006; Hamaus et al. 2009)
 \\
However the infrared and X-ray observations of \SGR are limited to one or two hour's durations of the flare events. Due to the short time durations, the reliability of the QPO periodicity is still open to argument.
Also the dependence of the detection upon observing instruments is another argument.(Meyer et al. 2008)

In contrast, because the \SGR is bright all time at radio wavelength,
 the continuous observations about 7 hours can be performed even from the northern hemisphere.
 The longer durations of radio observations of \SGR give us a chance to enhance the reliability of the QPO periodicity if the QPOs occur at radio wavelength. 
 Further, if we can observe the QPO periodicity of \SGR not only in intensity but also in structure change,
 the reliability of QPO detection will be conclusively enhanced.
In practice, however it has been recognized that the calibration of the VLBI data on \SGR is quite difficult
because of the atmospheric variations at the low elevation observations (Bower et al. 1999).
Then the VLBI synthesis imaging of \SGR is recognized as quite difficult, therefore the estimations of the size and shape of \SGR were performed through the closure analysis that are free from instrumental and atmospheric errors (Doeleman et al. 2001, Bowers et al. 2004, Shen et al. 2005)
As the closure quantities are not so robust against thermal noise,
 the use is limited to high SNR visibility data from shorter baselines (less than 2000 km).
Though the corresponding spatial resolutions or synthesized beam sizes are somewhat larger than the size of 
\SGR at 43GHz, the closure analysis can be performed successfully at the observing frequency.
Recently Miyoshi et al. (submitted to PASJ), by devising a simple method at calibration,
effectively used the data from longer baselines,
and achieved reliable synthesis imagings of \SGR at 43GHz, which are at least consistent to the previous closure estimations.
 Using the VLBA, we investigated the short time change with another new method, slit-modulation imaging method (Miyoshi 2008), and found four of periods with conspicuous patterned changes of the structure. 

\section{OBSERVATION AND DATA CALIBRATION}
 Our VLBA observations at 43 GHz were performed with $512$ Mbit-per-second recording rate ($16$ MHz $\times$ 8 IF channels, 2-bit sampling) from $9:30$ to $16:30 $(UT) on 2004 March 8$^{th}$, 
 37 $\sim$ 44 hours just after a millimeter wave flare of \SGR ~ detected with the Nobeyama Millimeter Array~(Miyazaki et al. 2006).

 In order to make efficient use of as much data as possible, we adopted a reduction strategy of keeping all data points till the final calibration stage where the CALIB (AIPS task of self-calibration) is performed. For data points where no solution was obtained at FRING in AIPS,
we assigned a value interpolated from adjacent good solutions to the data points.
Such keeping strategy in reduction is often performed for weaker maser data.
 We did the standard amplitude calibrations using antenna gains in GC table and system temperatures in TY table with opacity corrections in AIPS (NRAO). {Further at the last stage of the data calibrations we applied solutions of amplitude and phase from the task CALIB with an image model of the average size and shape of \SGR ~at $43$ GHz (single elliptical Gaussian with full width at half maximum of $712 \mu as \times 407 \mu as, PA = 79.8\DEG$, Bower et al. 2004) with $1$ Jy in flux density.
 The application of the CALIB solutions to the data} substantially reduced the effect of atmospheric and instrumental variations on the observed visibilities (Miyoshi et al. submitted to PASJ), and effectively made visibilities of longer baselines available for imaging synthesis.
 Figure 1 shows that the \UVC ~coverage of the solutions extends up to 5800 km ($8.3\times 10^8 \lambda$) in projected baseline length. This will give us a spatial resolution of three times better than those of the previous studies (Bower et al. 1998, 2004).
We show the amplitude and phase variations of the calibrated visibilities of respective baselines and closure phases for all triangles in Figures 2, 3, \& 4 and Figures 5, 6, \& 7 respectively.

%
\begin{figure*}
\begin{center}
\includegraphics[height=0.95\textwidth]{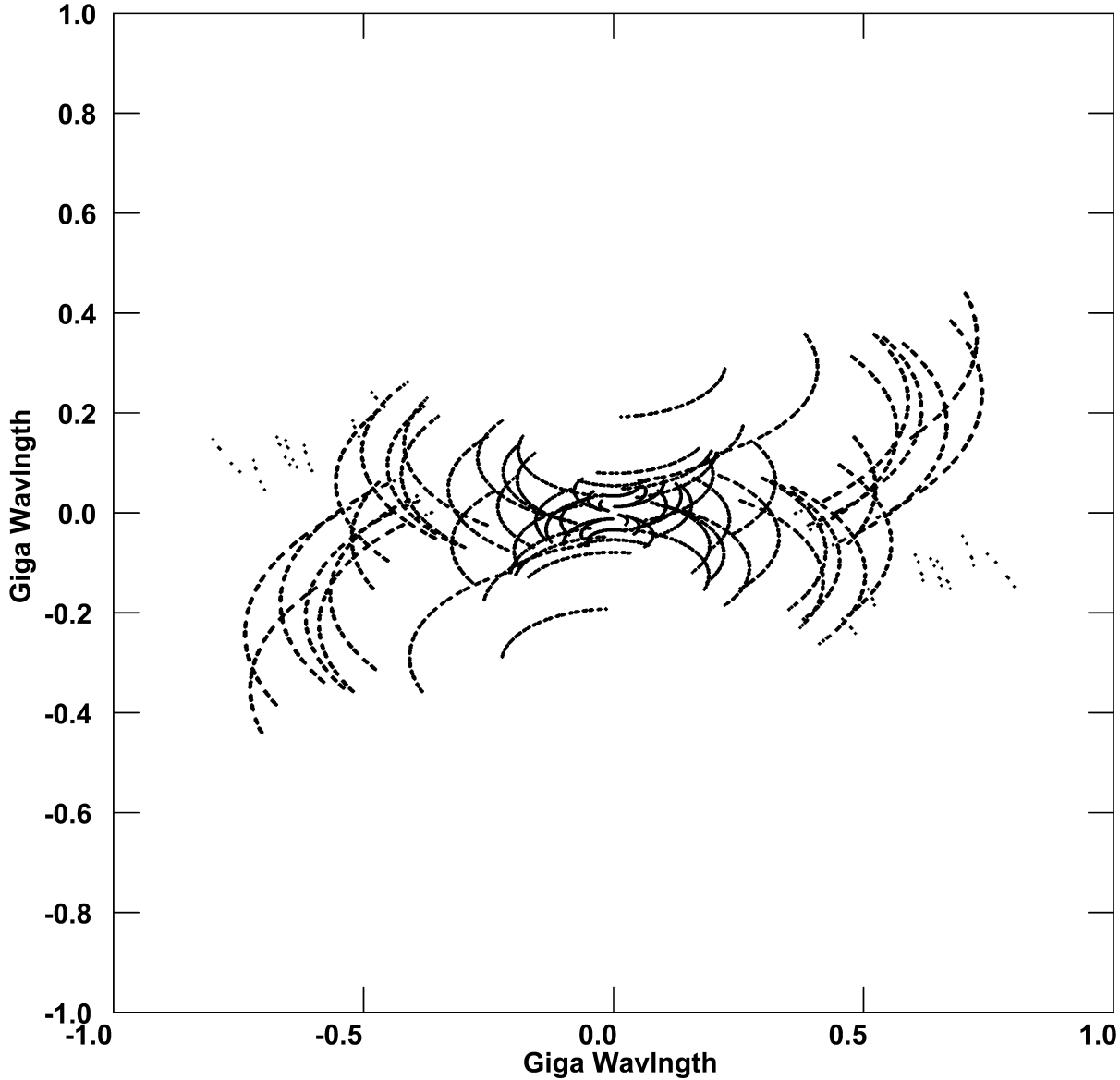} 
\caption{
 The \UVC ~coverage of all the visibility data that have calibration solutions by our reduction method.
 }
\label{UVPLT}
\end{center}
\end{figure*}
\begin{figure*}
\begin{center}
\includegraphics[height=1.225\textwidth]{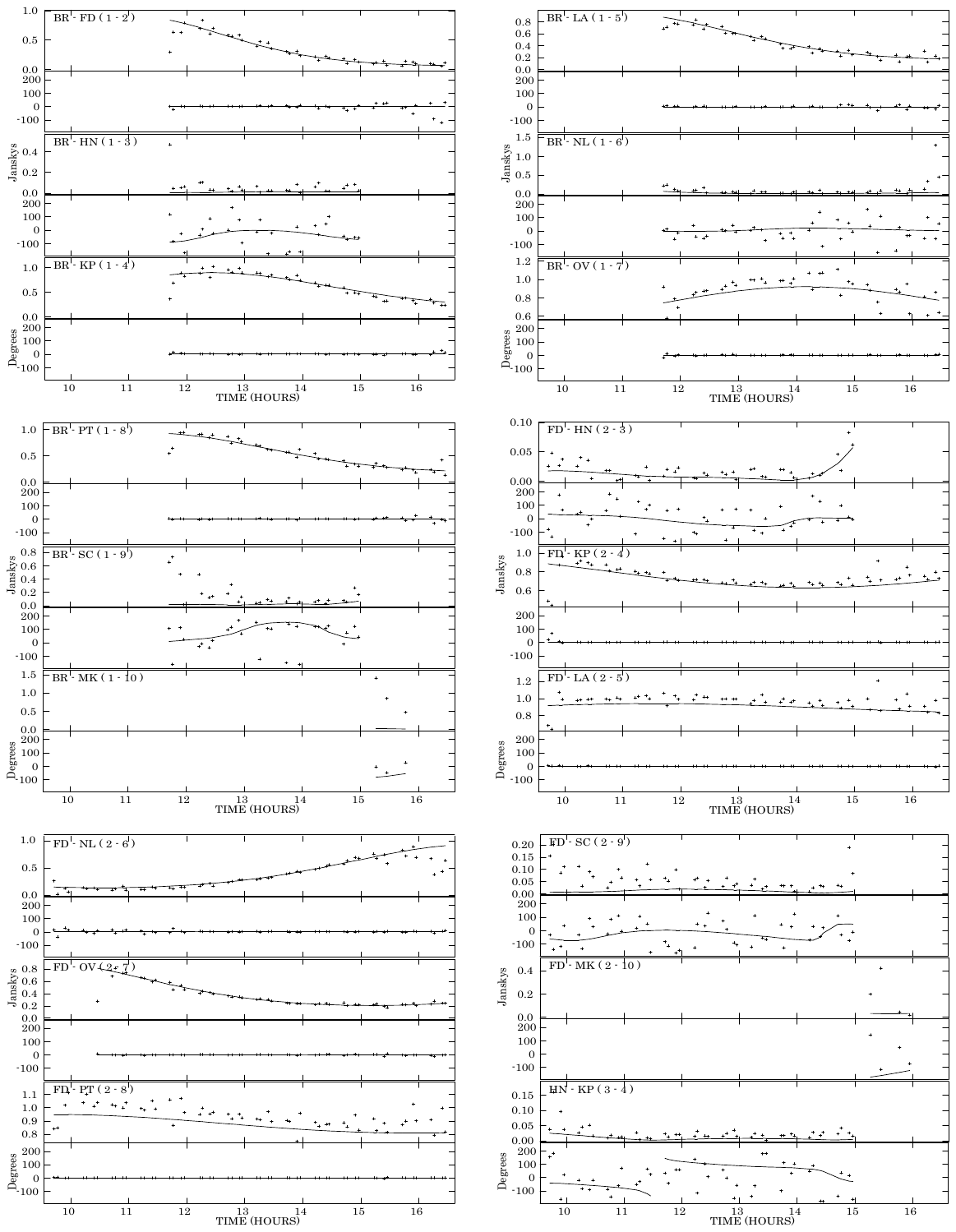}
\caption{ The time variations of visibility data in respective baselines (1/3). }
\label{fig VPLOT1}
\end{center}
\end{figure*}
\begin{figure*}
\begin{center}
\includegraphics[height=1.225\textwidth]{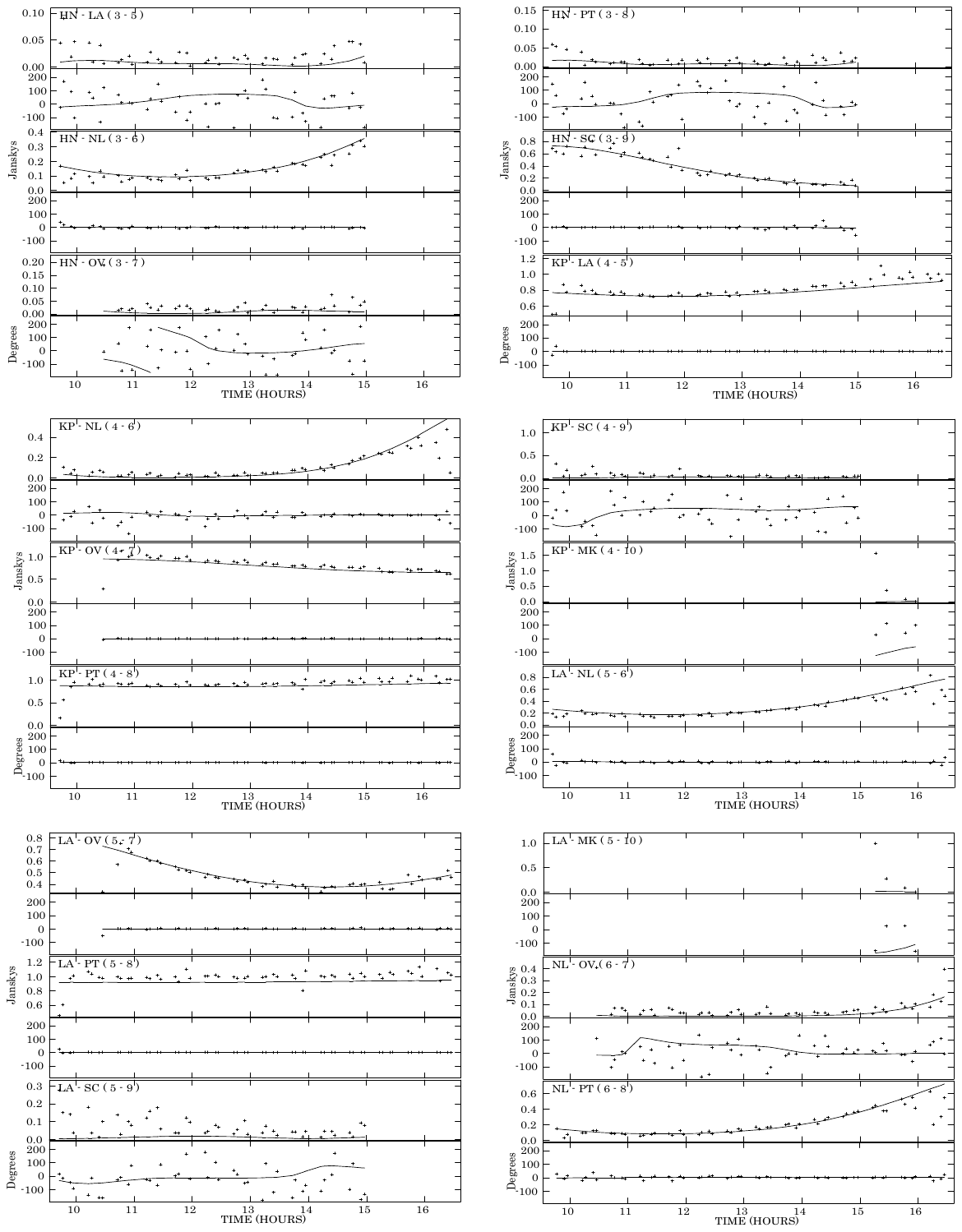}
\caption{The time variations of visibility data in respective baselines (2/3).}
\label{fig VPLOT2}
\end{center}
\end{figure*}
\begin{figure*}
\begin{center}
\includegraphics[height=1.225\textwidth]{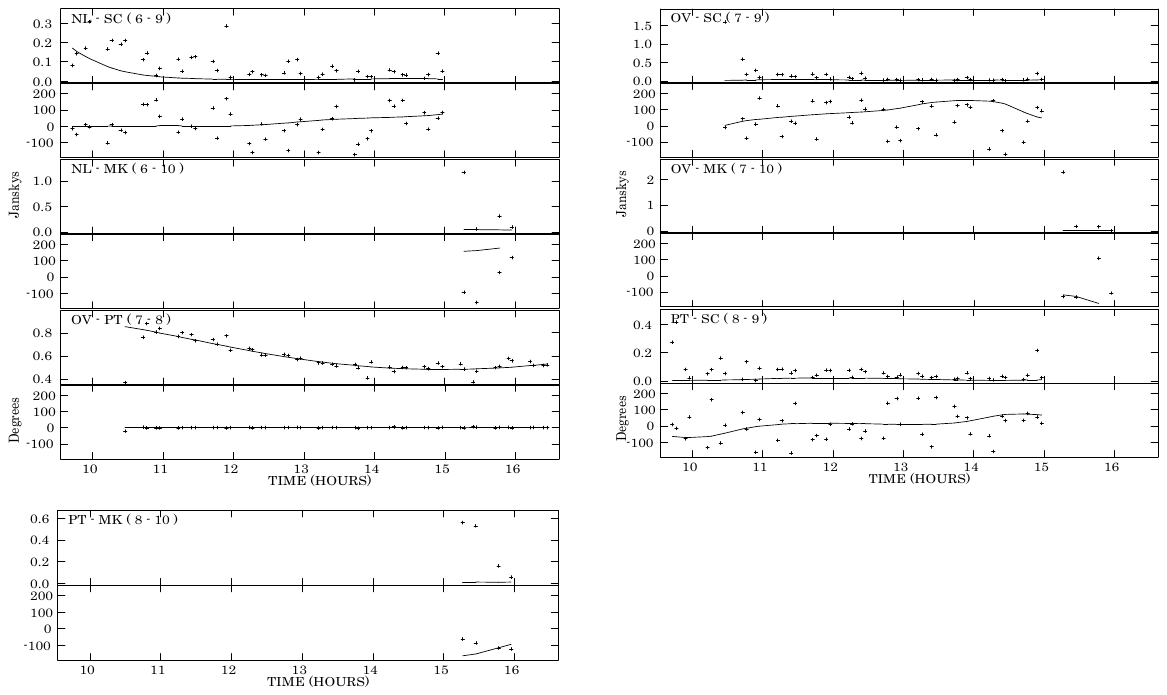}
\caption{The time variations of visibility data in respective baselines (3/3).}
\label{fig VPLOT3}
\end{center}
\end{figure*}

\subsection{The \UVC ~coverage of the calibrated data}
We got calibration solutions from 43 baselines out of all the 45 baselines, though not during the whole observing time.
While we found no solution from SC-MK and HN-MK baselines.
As for the baselines connected to MK, we found effective visibilities when the projected baseline length becomes shorter than 5800 km during the last one-half hours of the observing session. Except SC-MK and HN-MK,
we found solutions from the 7 baselines (BR-MK, FD-MK, KP-MK, LA-MK, NL-MK, OV-MK, and PT-MK).
As for the baselines connected to SC, we found solutions from the 8 baselines (BR-SC, FD-SC, HN-SC, KP-SC, LA-SC, NL-SC, OV-SC, and PT-SC). We found none of solution from the SC-MK baseline.
As for the baseline connected to HN, except the baseline between MK we found solutions from the 8 baselines (BR-HN, FD-HN, HN-KP, HN-LA, HN-NL, HN-OV, HN-PT, and HN-SC).
As for the baselines connected to BR, we found solutions from all the baselines.
Though the BR station locates at high latitude, we often find good visibilities of \SGR because the projected baseline length becomes shorter for the \SGR observations.
The all of resultant \UVC ~coverage distribute up to $8.3\times 10^8 \lambda$ (5800 km) in radius (Figure 1).
\subsection{Visibility variation in each baseline}
Figures 2, 3, \& 4 show the time variations of the amplitude and phase of each baseline
(each point is 5 min integration).
The lines in every plot show the time variations of the amplitude and phase calculated from the obtained
image from the whole time integration (Figure 8 a).
From the comparison between the real visibilities and calculated ones from the image,
we can classify them into 6 categories.
\begin{enumerate}

\item The phase is constantly zero and the amplitude has no variation with time.
Namely the source \SGR is observed as an unresolved point source
for the corresponding fringe spacing of the baselines:
2 baselines, KP-PT and LA-PT show this feature.
The maximum projected baseline lengths are about 400 km (KP-PT) and 210 km (LA-PT).
The corresponding fringe spacings are 3.5 mas and 6.7 mas respectively.

\item
The calculated phases and amplitudes from the obtained image, and the real ones
coincide each other well. The phases are constantly zero, while the amplitudes vary with
projected baseline length. The observed source \SGR is partially resolved by the spatial resolutions of the 
baselines. In other words, \SGR was observed to be not infinitely small but have a size:
19 baselines, namely BR-FD, BR-KP, BR-LA, BR-OV, BR-PT, FD-KP, FD-LA, FD-NL, FD-OV, FD-PT, HN-NL, 
HN-SC, KP-LA, KP-NL, KP-OV, LA-NL, LA-OV, NL-PT, \& OV-PT, belong to this category.
The average of these maximum projected baseline lengths is about 1400 km (1 mas in fringe spacing: from here in bracket after projected baseline length, we note the corresponding fringe spacing),
the shortest ones are about 560 km (2.5 mas) (FD-LA, FD-PT) and the longest one is about 2650 km (0.53 mas) (HN-SC).
\item
The calculated phases from the obtained image, and the real ones
coincide each other or show similarity. While the amplitudes from both do not show
similarity. 3 baselines, BR-SC, BR-MK, PT-MK, belong to this category.
The maximum projected baseline lengths are 5740 km (0.24 mas), 3850 km (0.36 mas), and 4760 km (0.29 mas) respectively.

\item
 The calculated phases from the obtained image, and the real ones
do not coincide each other. While the amplitudes from both show similarity and match
in some degree:
12 baselines, BR-HN, BR-NL, FD-HN, HN-KP, HN-LA, HN-OV, HN-PT, KP-SC, NL-OV, HN-SC, OV-SC, \&
PT-SC, belong to this category.
The average of these maximum projected baseline lengths is about 3710 km (0.38 mas),
the shortest one is about 2380 km (0.59 mas) (NL-OV) and the longest one is about 5530 km (0.25 mas) (OV-SC).
\item
The calculated visibility and the real ones do not show matching both in phase and in amplitude:
7 baselines, FD-SC, FD-MK, KP-MK, LA-SC, LA-MK, NL-MK, \& OV-MK, belong to this category.
In general there are at least three possibilities to explain the no-matching.
One is due to a low signal to noise ratio including the case of no signal.
Another is due to the complex structure of the real source, the obtained rough image could not reproduce the
real visibility variations. The other is the mixture of the two cases above.
The average of these maximum projected baseline lengths is about 4650 km (0.30 mas),
the shortest ones are about 3920 km (0.36 mas) (NL-OV) and
 the longest one is about 5670 km (0.25 mas) (OV-SC).
\item
No solution is found in SC-MK and HN-MK baselines, which are longer than 5800 km in projected baseline length.
\end{enumerate}

Assuming that matching between the observed and the calculated visibilities
mean sufficient calibrations of the two stations of the baseline,
we can guess whether the calibrations both of the stations are good or bad.
(Here we think that mis-matching does not necessarily mean bad calibrations.
If the observed visibility is from time variable source in structure,
 any single image cannot satisfy the whole time visibility variation.
 However, if the structure variation is not so large one, such a small variation is negligible and not detectable for shorter baselines.)
By eye, we found that the following baselines seem in the matching:
BR-FD, BR-KP, BR-LA, BR-OV, BR-PT, FD-KP, FD-LA, FD-NL, FD-OV, FD-PT, HN-NL, HN-SC, KP-NL, KP-OV, KP-PT,
 LA-NL, LA-OV, LA-PT, NL-PT, and OV-PT.
If the assumption is correct, except the MK station, other 9 stations seem to be well-calibrated anyway.

\subsection{Closure phases}
Figures 5, 6, \& 7 show the time variations of the closure phases of all triangles composed from three stations. Every point is from 15 min integration.
We also calculated the closure phase variations from the images of the whole time integration as shown in Figure 8 (a). Closure phases show several features as below.

\begin{enumerate}
\item 
First feature is seen in the small triangles composed from inner stations of the VLBA, namely FD, LA, PT, KP, \& OV. The projected extensions of the triangles range from 700 km (2 mas) to 1600 km (0.88 mas), and the average is 1200 km (1.2 mas). These 10 triangles show nearly zero constant closure phases, a few degree variations at most. This indicates that \SGR is observed as a structure of point symmetric with the corresponding fringe spacings ($0.88 - 2 mas$).
\item
Second feature is seen in the 10 triangles composed from BR and other two stations from the inner 5 stations, FD, LA, PT, KP, \& OV. The projected extensions of the triangles range from 1500 km (0.93 mas) to 2200 km (0.64 mas), and the average is 1900 km (0.74 mas). The closure phases show zero on average, but fluctuating $\pm 40 \DEG$ at peak.
\item
Third feature is seen in the 10 triangles composed from NL and other two stations from the inner 5 stations, FD, LA, PT, KP, \& OV. The projected extensions of the triangles range from 1800 km (0.78 mas) to 2800 km (0.5 mas), and the average is 2400 km (0.58 mas). The closure phases show zero on average, but fluctuating by $\pm 60 \DEG $ on average, $\pm 100 \DEG $ at maximum case.
\item
In most of large triangles which have projected extensions up to 5800 km (0.24 mas),
the closure phases distribute $\pm 180 \DEG$.
However, some of triangles including MK station show nearly constant and/or 
very small variations in closure phase though the duration is about 40 minutes.
Two triangle of FD-NL-MK and LA-PT-MK show closure phase variation within $15 \DEG$,
triangle of KP-PT-MK shows that within $10 \DEG$, and
triangle of KP-LA-MK shows nearly constant closure phase within $8 \DEG$ variation.
The projected extensions of these three cases range from 4800 km (0.29 mas) to 5200 km (0.27 mas), 4900 km (0.29 mas) respectively.

\item
In most of the large triangles, the real closure phase and the calculated phase from the
obtained image do not match. While one of the large triangles, HN-NL-OV shows matching between them.
The projected extension of the triangles is about 4200 km (0.33mas).

\item
The last feature is seen in the five of triangles BR-FD-KP, BR-FD-LA, BR-FD-NL, BR-FD-OV, BR-FD-PT, 
whose projected extensions range from 2100 (0.67 mas) km to 2500 km (0.56 mas), and the average is 2300 km (0.64 mas). They show very similar closure phase variations each other.

\end{enumerate}

\begin{figure*}
\begin{center}
\includegraphics[height=1.225\textwidth]{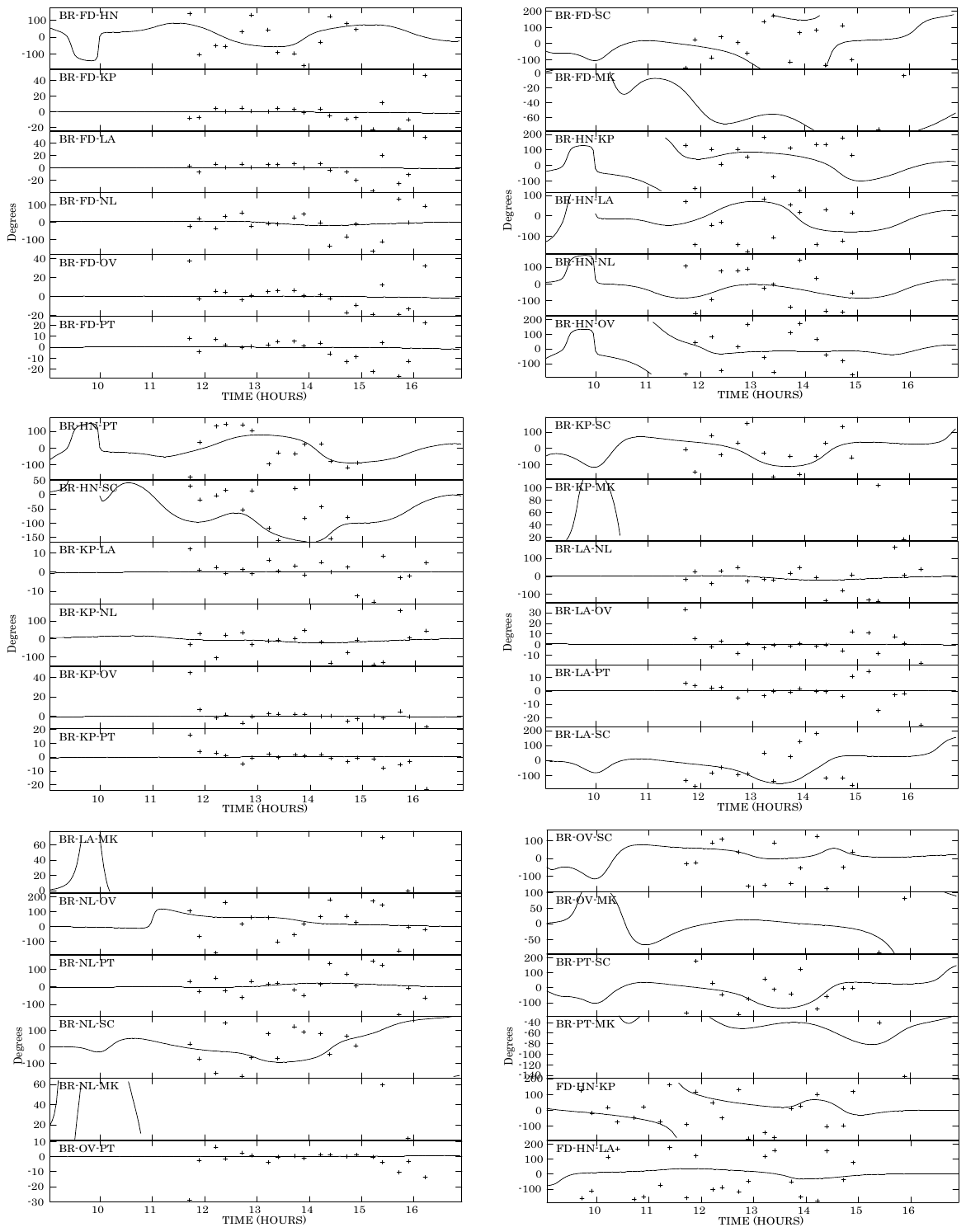} 
\caption{
 The time variations of closure phases in respective triangles (1/3).
}
\label{fig CLPLT1}
\end{center}
\end{figure*}
\begin{figure*}
\begin{center}
\includegraphics[height=1.225\textwidth]{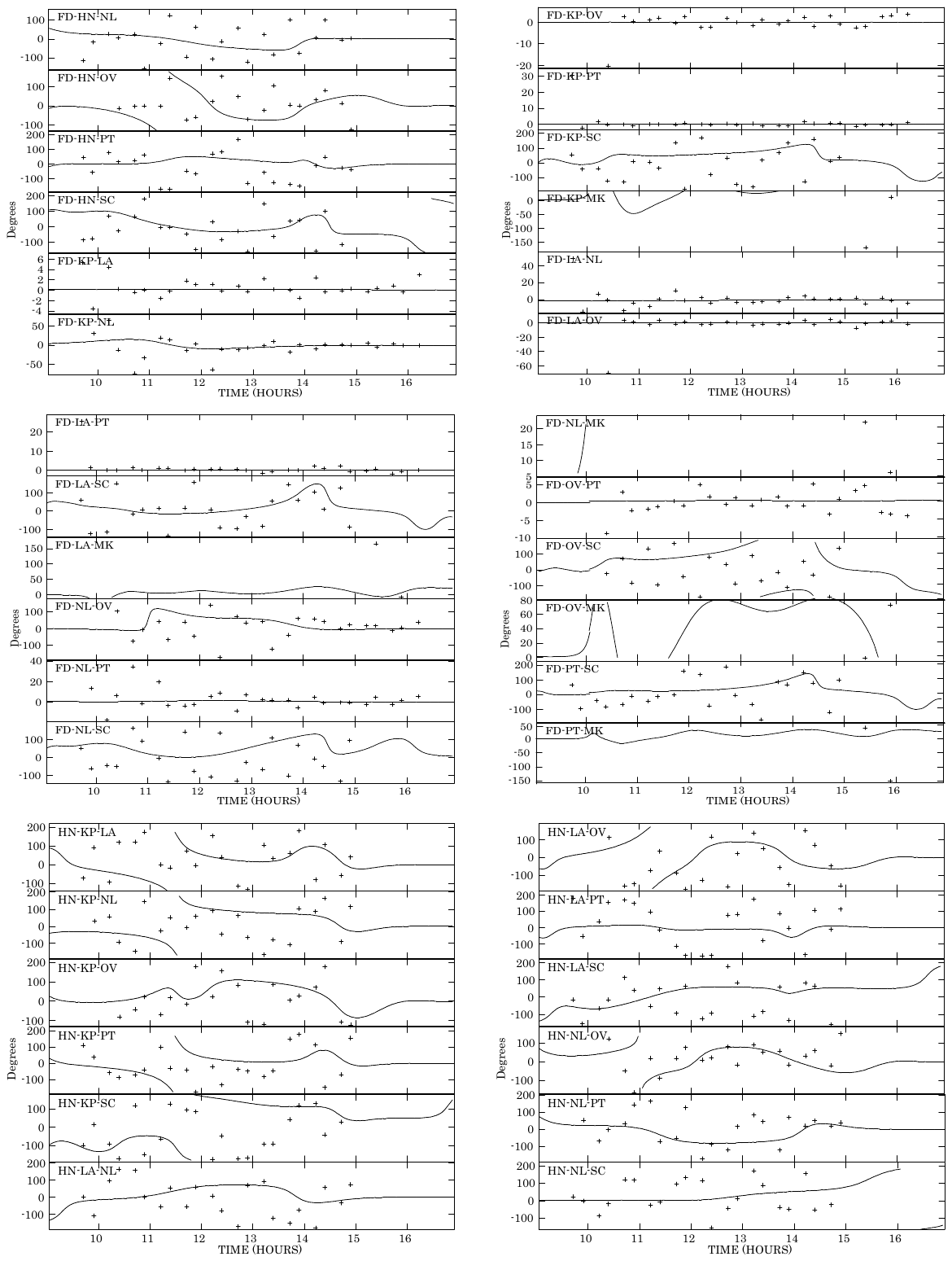} 
\caption{
 The time variations of closure phases in respective triangles (2/3).
 }
\label{fig CLPLT2}
\end{center}
\end{figure*}
\begin{figure*}
\begin{center}
\includegraphics[height=1.225\textwidth]{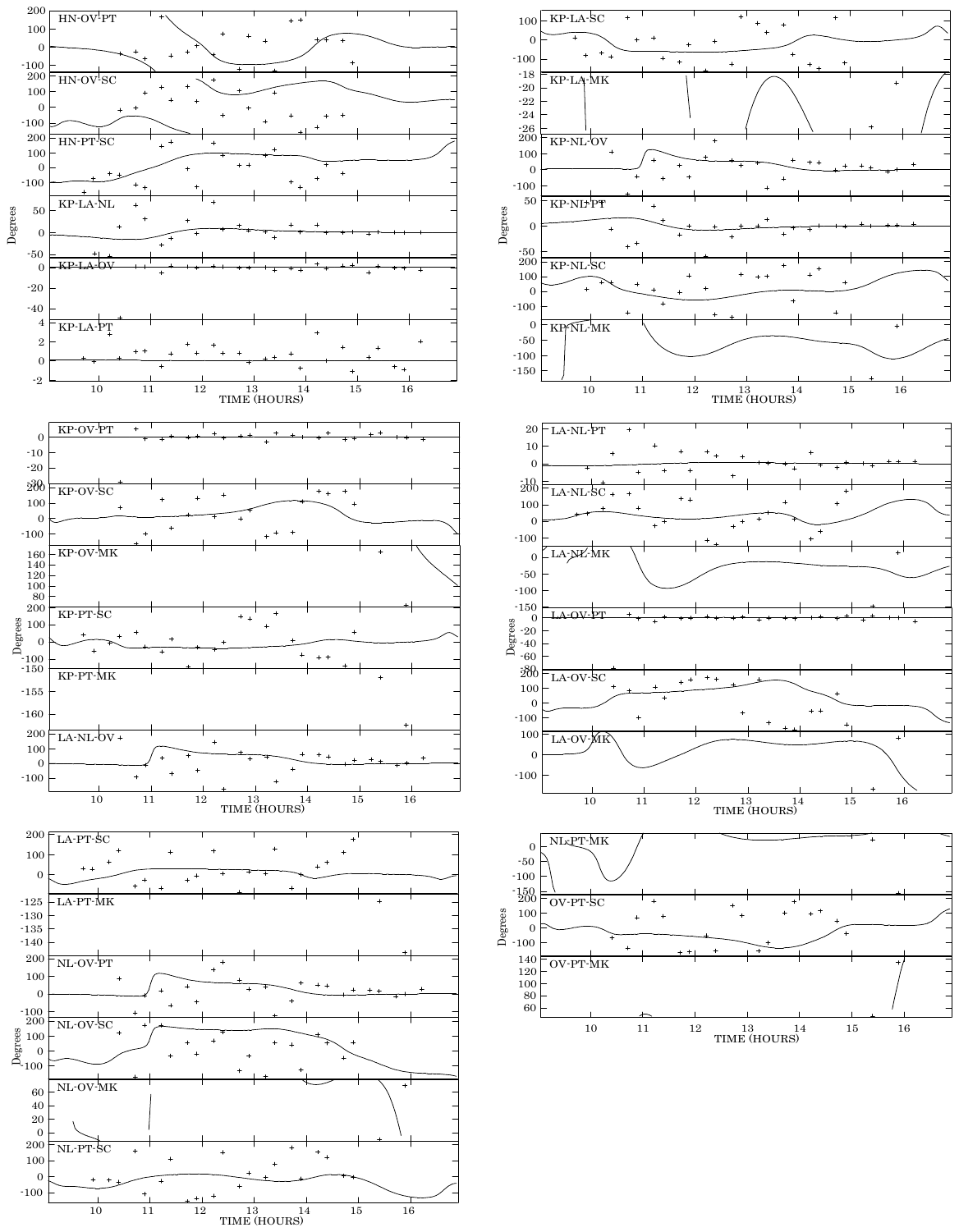} 
\caption{
 The time variations of closure phases in respective triangles (3/3).
 }
\label{fig CLPLT3}
\end{center}
\end{figure*}

\subsection{The calibrations of MK, SC, and HN stations}
Visibility plots (Figure 2,3, and 4) and closure plots (Figure 5,6, and 7) show that the SC and HN stations are calibrated, though not perfect. The visibility plots of HN-NL, and HN-SC shows quite a good match between the real amplitudes \& phases and the calculated ones from the obtained map. The visibility plot of BR-SC shows good match in phase variation, that of FD-HN shows good match in amplitude variation. Presumably, the HN and SC stations are calibrated quite well.
At least, the calibrations to the HN and SC stations remove largely the systematic errors.\\
About the MK station, the calibration solutions were obtained for the last one hour. Because the duration is short, the match in visibility is difficult to confirm.
 But, some of the triangles including MK station show nearly constant and/or 
very small variations in closure phase.
These features presumably mean that the MK station got fringes between other stations though the signals are weak. \\
 In APPENDIX D, we show several images by omitting the visibility data of the baselines to the MK, SC, HN stations. These images suggest that the calibration errors and the low SNR of the data from three stations have little influence upon the image quality.

\subsection{Obtained image from the whole time data integration}
We made an image from the whole time integration using the task IMAGR in AIPS with loop gain parameters (GAIN) = 0.005, limit of subtracting iterations (NITER) = 20000, maximum residual flux density level (FLUX) = 0.005 Jy, and \UVC ~Gaussian tapering (UVTAPER) = 6000 km. Here we used the wider tapering in \UVC ~with expectations of being able to use lower SNR data points by the CALIB solutions.
The mapping area is $ 3~mas~(e-w) \times 6~mas~(n-s)$ and the boxing area is $ 2.13~mas~(e-w) \times 4.17~mas~(n-s)$ at the center. Because the sizes of \SGR at 43GHz are well known, we limited the peak search to the central area to avoid selecting side lobe area.
The grid numbers are $1024~(e-w) \times 2048~(n-s)$, namely the grid spacing is nearly $2.93\mu as$.
By using a smaller grid, we intended to select the peak positions as accurately as possible. 
 We performed the imaging with a restoring beam of $0.40~mas \times~0.15~mas, PA = 0 \DEG $\\
 The obtained image from the whole time integration suggests that the calibration of the visibilities
are successfully performed.
From direct fitting of an elliptical Gaussian shape 
to the image and the deconvolution of the restoring beam shape,
we got a result of the shape and size. 
The major axis is $0.759^{+0.006}_{-0.007}$ mas, 
the minor axis is $0.323\pm0.007$ mas, and 
the position angle is $PA=85.1\pm 1^{\circ}$.
This estimation is consistent to the previous measurement of the same data using closure amplitude
by Shen et al. (2005) that show the major axis is $0.722\pm0.002$ mas, and the minor axis is
$0.395^{+0.019}_{-0.020}$ mas and the PA is $80.4\pm0.8$ $^{\circ}$.
The deviations from the Shen's measurement are
$ +0.037$ mas in major axis, $-0.072$ mas in minor axis and $4.6 ^{\circ}$ in PA.
 Miyoshi et al. (submitted to PASJ) show the details of the results from other VLBA data on \SGR
 at 43GHz with the same reduction manner.
The consistent Gaussian shape and size to those found Shen et al. (2005) indicates that we got better calibration solutions from the CALIB and that we successfully obtained images of \SGR 
with better spatial resolution than before.
%

\begin{figure*}
\begin{center}
\includegraphics[height=1.05\textwidth]{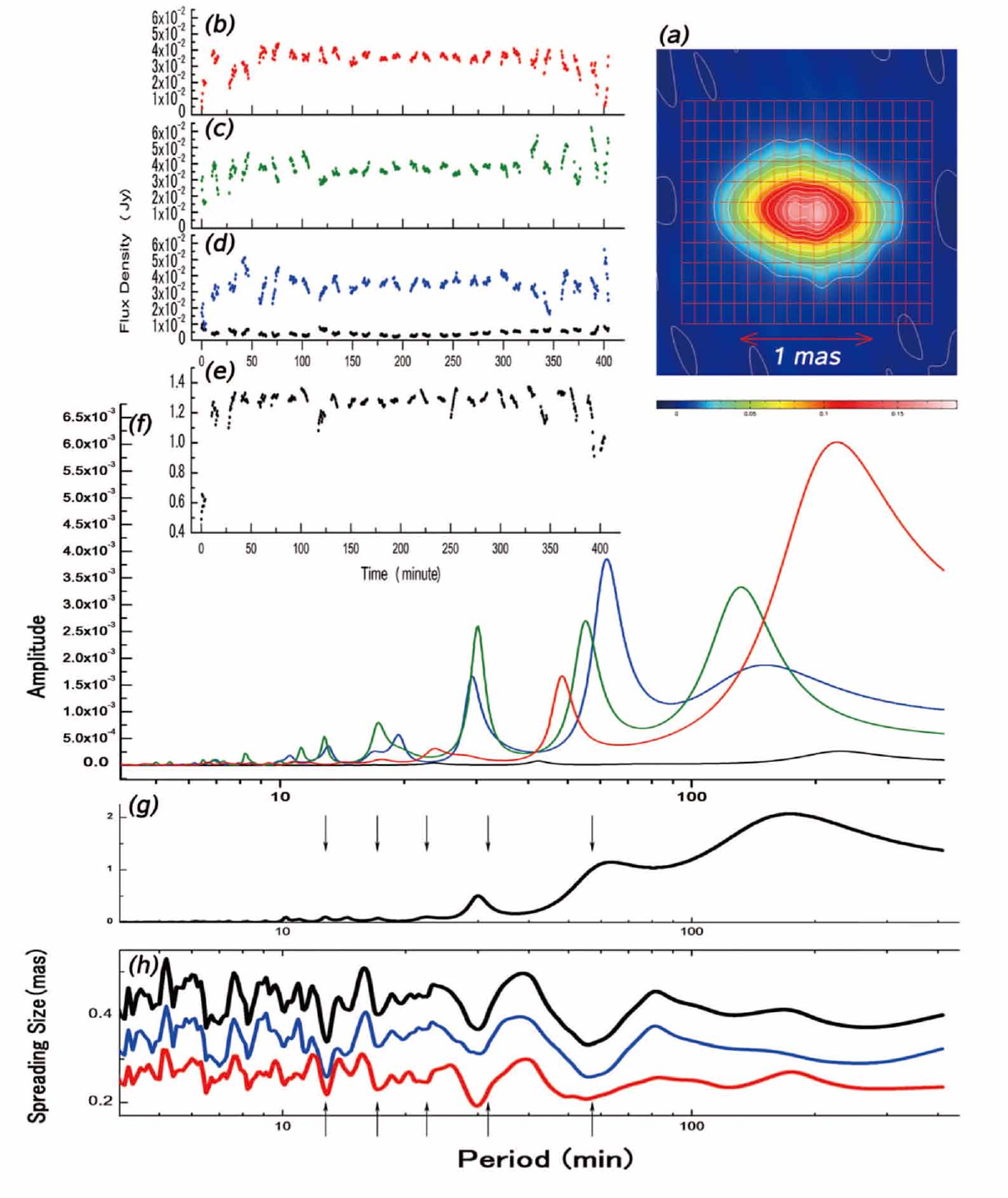}
\caption{
The top right panel (a) shows the Sgr A$^{*}$ image from the whole observing time. 
The contour levels are every 10 \PC~of the peak brightness.
We defined 209 small rectangles of 0.1 mas in east-west and 0.15 mas in north-south around the map center, distributed as matrix 19 in east-west by 11 in north-south.
We measured the time variations of flux densities of the rectangles.
 Panels (b), (c), and (d) show the flux density variations of the central three rectangles, 0.1 mas west, center, 
 and 0.1 mas east along the central east-west line respectively. 
 Panel (e) shows the time variations of the total of 209 rectangles.
 The rms noise levels, measured at the outer region of the maps where no real emission should come are also shown as black points in Panel (d).
 The intensity integrated over 209 rectangles is about 1.3 Jy, which is the typical flux density of SgrA$^{*}$ at 43 GHz,
 and those of the central 3 rectangles are 7 times higher than that of the rms noise level.
 Panel (f) shows the power spectra of periodicity calculated from the time variations using MEM.
 The x-axis is the period (minute) and the y-axis is the amplitudes (arbitrary unit). The blue line is that of the 0.1 mas-east rectangle,
 the green line is that of the central rectangle, and the red line is that of the 0.1 mas-west rectangle.
Panel (g) shows the spectra of the flux density variation of all 209 rectangles.
 The black line in Panel (f) means the power spectra of the noise variations shown in Panel (d). 
The peaks of noise spectra appear at different periods from those of peaks of time variations in the central three rectangles.
The bottom panel (h) shows the spreading size, one standard deviation of the distribution of each period of the area of the 209 rectangles.
The black line shows the absolute size, while the blue line shows the component of the east-west direction.
 The red line shows that of the south-north direction.
 The five arrows show the positions of $P = 12.9, 16.8, 22.2, 31.4, \& 56.4$ min from right to left respectively.
 At these periods, except $P = 12.9$ min, we found periodic structure changes from SMI method.}
\label{fig 1}
\end{center}
\end{figure*}

\section{ANALYZING METHOD AND RESULTS}

\begin{figure*}
\begin{center} 
\includegraphics[height=1.05\textwidth]{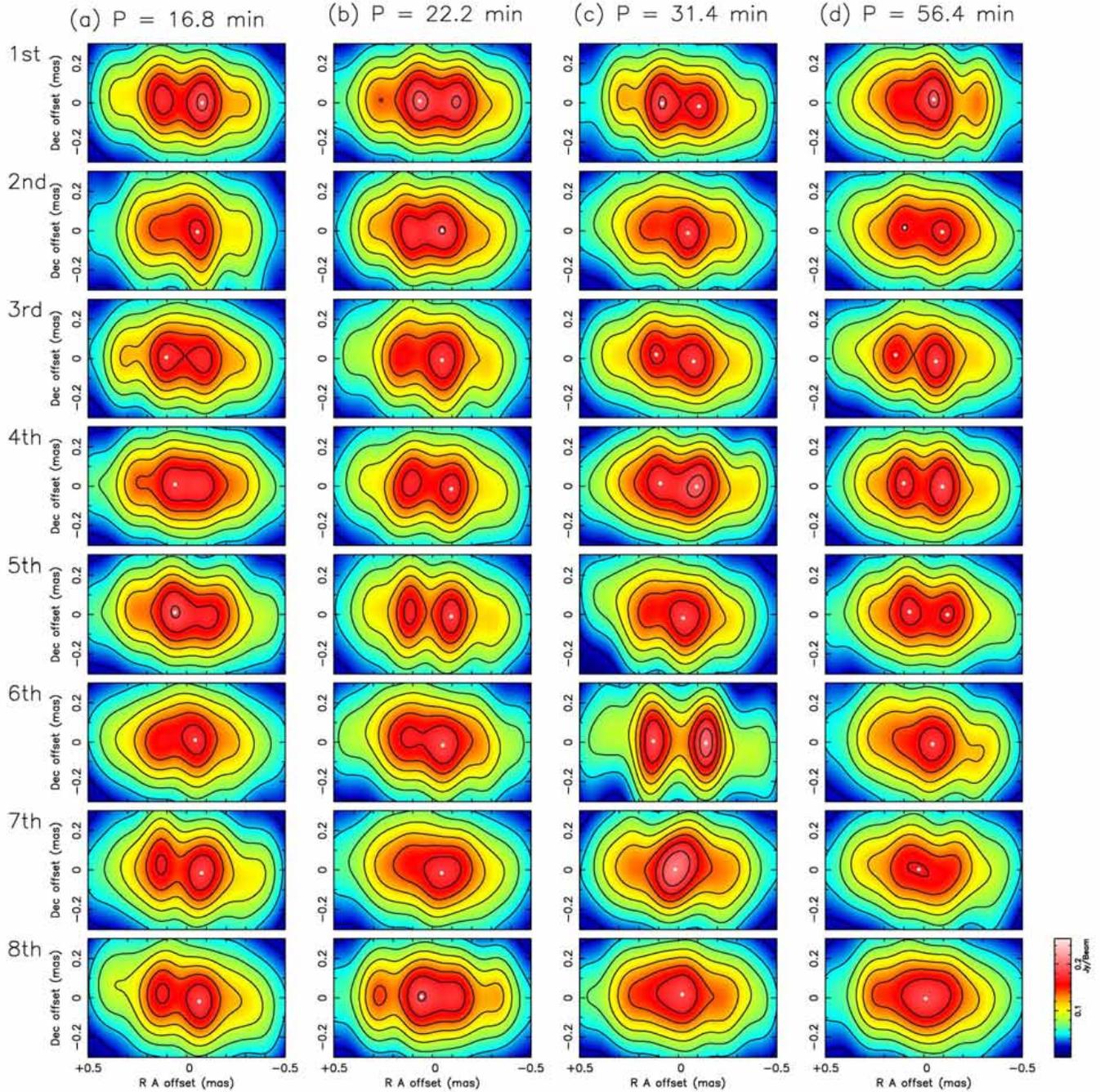} 
\caption{
 The slit-modulation-imaging~(SMI) maps with the 4 periods $P = 16.8 (a), 22.2 (b),$ $31.4 (c), and ~56.4$ min (d). The 8 maps show the images in phase frame series with respective periods.
 The white filled circles mark the intensity peak position in each map. In $P = 31.4~\&~56.4$ min,
second peaks are also marked with circles. The contour levels are every 10 \PC~of the peak brightness of the respective maps.}
\label{fig 2}
\end{center}
\end{figure*}

We investigated the spatial distributions of oscillations in flux density with two independent methods. The first one is the comparisons of power spectra of time variations of intensities in 209 small rectangles areas of the image (Figure 8 a). We produced 360 snapshot maps from 5-minute integrations with a starting time increment of 30 seconds. 
 Each map was produced by the same parameters as explained in the previous section except the maximum residual flux density level (FLUX) =10mJy.
 
 We then measured flux density variation in each small rectangle.
Their time variations of the flux densities show small up-and-downs similar to, but not the same as, those of other rectangles (Figure 8 b, c, d, and e).
 We then calculated their power spectra using a maximum entropy method (MEM, Burg 1967).
In Figure 8 (f) are shown the power spectra of the central 3 rectangles, whose SNRs are more than 7,
while in Figure 8 (g) are the spectra of time-variations of all 209 rectangles shown in Figure 8 (e).

 Figure 8 (h) shows the spreading size $\sigma(P)$, namely the one standard deviation of the spatial distribution of the amplitude of each period.
The definition is
\begin{equation}
 \sigma(P)  =sqrt[ \sum_{i=1}^{209} (x_{i}^2+y_{i}^2) \cdot A_{i}(P)/S(P) ]
\end{equation}
The component in x direction is
\begin{equation}
 \sigma_{x}(P)=sqrt[ \sum_{i=1}^{209} x_{i}^2     \cdot A_{i}(P)/S(P) ]
\end{equation}
The component in y direction is 
\begin{equation}
\sigma_{y}(P)=sqrt[ \sum_{i=1}^{209}     y_{i}^2 \cdot A_{i}(P)/S(P) ]
\end{equation}

,where $S(P)=\sum_{i=1}^{209} A_{i}(P)$, the sum of the amplitudes in the 209 rectangles.
 The i means $i^{th}$ rectangle, P is the period, $A_{i}(P)$ is the amplitude of the period in each rectangle, ($x_{i}$, $y_{i}$) is the central position of the $i^{th}$ rectangle.
If all of the power of the period concentrate at the central rectangle $i=0$, the spreading sizes
 $\sigma(P)$, $\sigma_{x}(P)$, \& $\sigma_{x}(P)$ are zero.
 If the power distributes uniformly in all the rectangles, the spreading sizes are
0.362 mas in $\sigma(P)$, 0.274 mas in $\sigma_{x}(P)$, and 0.237 mas in $\sigma_{x}(P)$.
In the case that the spreading sizes are larger than these values, 
the power of the period mainly comes from outer area, where the flux density and the SNRs are lower. Such periods with larger spreading sizes presumably originate from
 noise.
If the spreading sizes of the periods are smaller than these values,
the power of the periods comes from the central part of the image, 
where the flux density and the SNR is higher.
A real periodicity, if detected, because it should come from the central part of the image,
the spreading size should be smaller than that of the uniform case.

Small spreading sizes, or the concentrations of the distributions at the center, are seen around $P = 12.9, 17.2, 30.1, and ~55.8$ min, where the amplitude shows a maximum (Figure 8 g). 
Loose concentrations appear around $P = 128.4, and ~268$ min though they are under-sampled.

\begin{figure*}
\begin{center}
\includegraphics[height=1.25\textwidth]{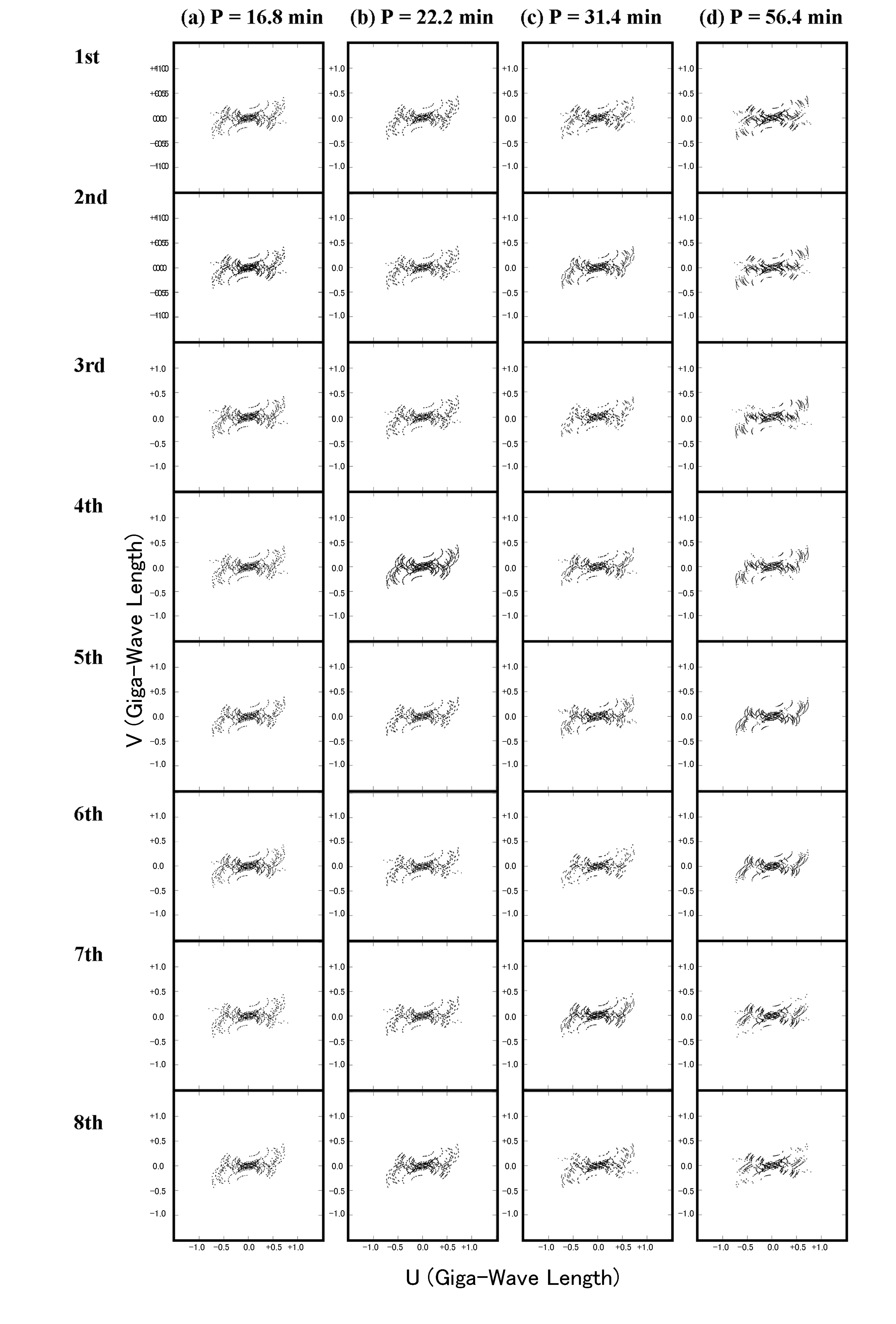}
\caption{The \UVC ~coverage of the respective SMI maps.}
\label{UVC-SMI}
\end{center}
\end{figure*}
 The first method tends to be affected by differences of u-v sampling between snap shot maps.
 We hence devised and used a new method, slit-modulation-imaging (SMI) which is a way of examining the existence of the periodic structural change with an assuming period and almost free from u-v sampling differences~(Miyoshi, 2008). SMI maps are obtained as follows: we divided the whole observation time into several sections with the trial period P min interval and then each section was sub-divided into 8 phase-segments. The first-phase SMI map was produced with visibilities from the first phase-segments of all the sections. The second-phase SMI map was produced with visibilities from the second phase-segments of all the sections. Namely, the n-th phase SMI map is produced from the visibilities of the n-th phase-segments of all the sections. Thus, the SMI maps have \UVC ~ ~coverage very similar to each other, and are almost free from the effect of differing \UVC ~coverages (Figure 10). Characteristic structure changes with the period will be enhanced and emerge onto the respective SMI maps if it exists.~(Miyoshi 2008)
 
 We tested around these confined periods of $P = 12.9, 17.2, 30.1$, and $55.8$ min, adding with the $P = 22.2$ min detected at X~ray flares~(Belanger et al. 2006, Hamaus et al. 2009) (Figure 8 (f) also shows a small peak at the period).
 Except for $P = 12.9$ min, we found conspicuous periodic change patterns from 
 $P = 16.8, 22.2, 31.4$, and $56.4$ min, with estimated errors $\pm 1.4, \pm 1.4, \pm 1.5, \pm 6 $ min respectively. Because of the findings of simple change patterns, we regard the 4 periods as confirmed detection. They are very roughly commensurable in a $3:4:6:10$ ratio.
\begin{figure*}
\begin{center}
%
\includegraphics[height=0.8\textwidth]{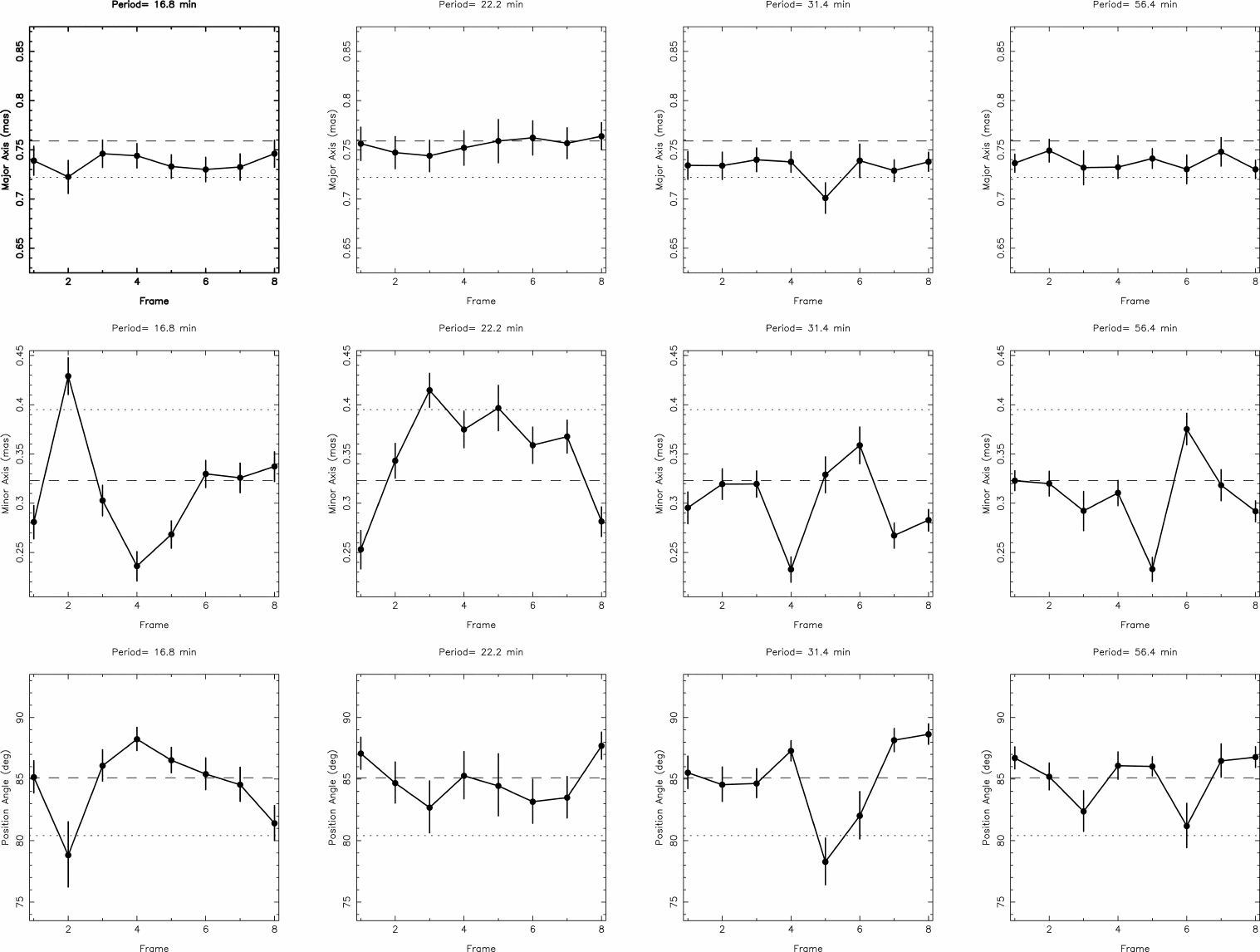} 
\caption{ Gaussian size estimations of the each SMI maps. The dashed line shows the parameter
obtained from the whole time integration image. The dot line shows the parameters
estimated from closure analysis (Shen et al. 2005).}
\label{fig SMIGAUS}
\end{center}
\end{figure*}
%
 At first glance these SMI results seem in stark contrast with the previous VLBA observations of \SGR at 43GHz. For quantitative evaluation of the difference, we performed an elliptical Gaussian fitting directly to the obtained SMI maps, and the results of deconvolution of the restoring beam show that these SMI maps,
if we regard them as elliptical Gaussian shapes, are consistent to the previous size \& shape estimation 
of \SGR from closure analysis (Figure 11). The differences are mostly within error bars of Bower's estimations (Bower et al. 2004) but out of the small error bars of Shen's estimations (Shen et al. 2005)

 In the case of $P = 16.8$ min, the peak is in the west of the image at the $1^{st}$ and the $2^{nd}$ phase frames, but shifts eastward, stays in the east at the $3^{rd}$, $4^{th}$ and $5^{th}$ phase frames, and after that, moves westward again (Figure 9 a). \\
In the case of $P = 22.2$ min, at the $1^{st}$ phase frame the peak is in the east, moves towards the west at the $2^{nd}$ phase frame, 
shifts further west at the $4^{th}$ and $5^{th}$ phase frames, and then moves back towards the east from the $6^{th}$ to $8^{th}$ phase frames (Figure 9 b).
Namely, in the $P = 16.8$ and $22.2$ min structure changes, the brightest position oscillates mostly in the east-west direction with the respective periods, which looks like a one-arm structure ({\it m} = 1) motion viewed edge-on.\\
 Assuming the motions to be projected circular ones, we found
the $P = 16.8$ min orbit is $0.18$ mas while that of the $P = 22.2$ min is $0.15$ mas in diameter, both are tilted with $PA = 82\DEG$ nearly the same position angle as the major axis of the average \SGR image at 43 GHz (Figure 12). The viewing angles are almost edge-on, $87.0\DEG$ ($P = 16.8$min) and $87.7\DEG$ ($P = 22.2$ min) respectively (Table 1).\\

\begin{figure*}
\begin{center}
\includegraphics[height=0.8\textwidth]{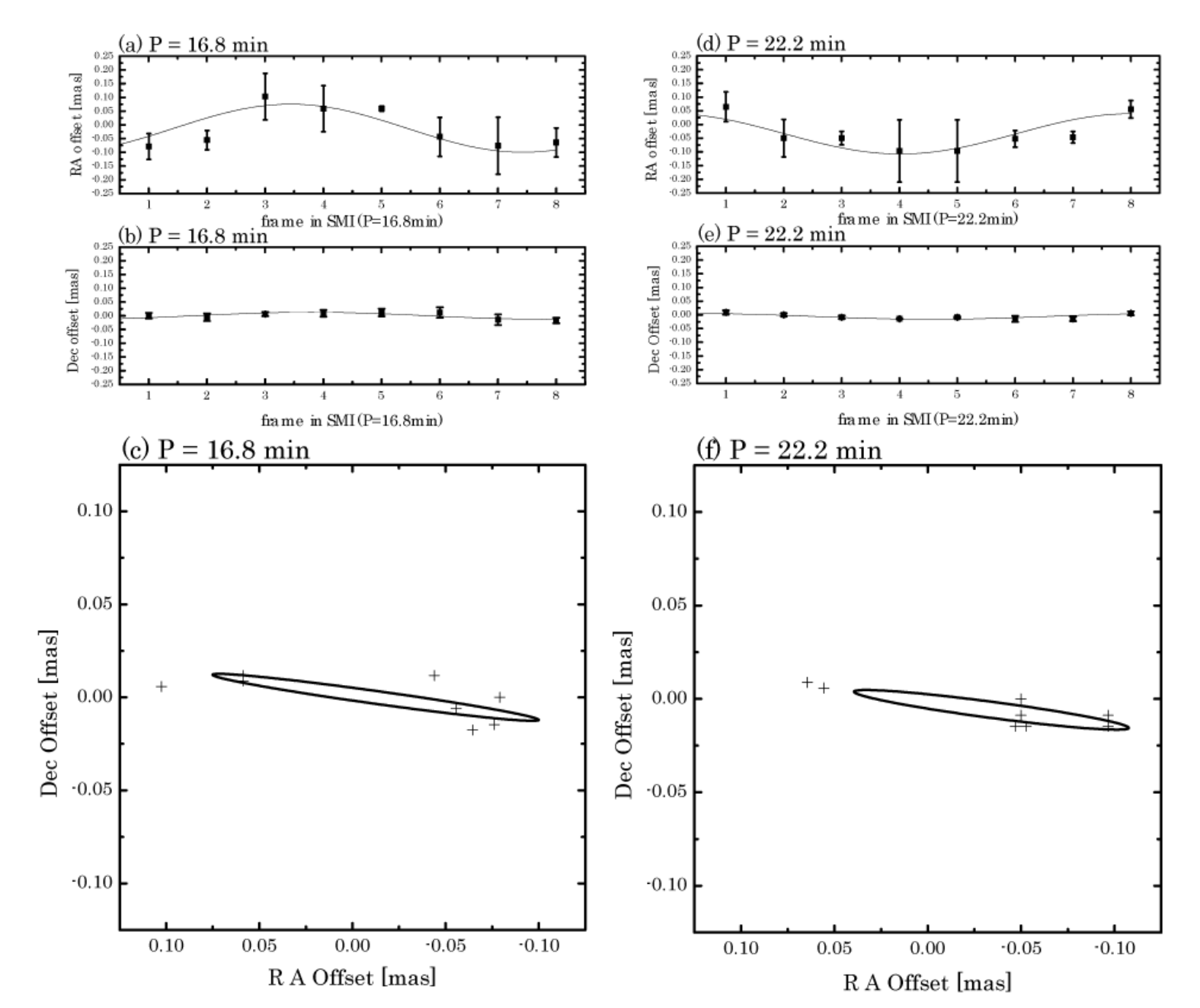}
\caption{The offset of brightness peak positions in the frame series of SMI images: those of P= 16.8 min in the left panels and those of P= 22.2 min in the right panels.
 The top panels show the positional shifts in R.A. while the middle panels in declination. The curve in each panel shows the sinusoidal function obtained from fitting to the data.
The bottom panels show the displacements of the peaks in the sky. The ellipses are formed from the obtained fitted sine curves.}
\label{fig motion}
\end{center}
\end{figure*}

\begin{table}
\begin{center}
\begin{tabular}{|c|cc|c|c|c|}
\hline
Period & \multicolumn{1}{|r}{Center} & \multicolumn{1}{l|}{Position} & Diameter & PA & Inclination \\
 & RA & DEC & & & \\
(min) & (mas) & (mas) & (mas) & ($\DEG$) & ($\DEG$) \\
\hline
16.8 & -0.02 & 0.00 & 0.177 & 81.8 & 87.7 \\
\hline
22.2 & -0.03 & -0.01 & 0.149 & 81.8 & 87.0 \\
\hline
\end{tabular}
\caption{Fitted parameters of circles
to the displacements of the brightest peaks in the P = 16.8 \& 22.2 min}
\label{}
\end{center}
\end{table}

In the $P = 31.4$ min change, one or two peaks appear in each phase frames (Figure 9 c). The numbers of peaks are $2\rightarrow 1\rightarrow 2\rightarrow 2\rightarrow 1\rightarrow 2\rightarrow 1\rightarrow 1$ with phase frame change.
At first glance this seems random, but a relation exists.
 The sum of peak numbers in the two SMI maps just half a period away from each other is always three, which can be explained by the existence of a three-arm pattern ({\it m} = 3) with evenly separation angle rotating with the period, which we view edge-on. \\

 In the P = 56.4 min change, one peak appears during a half period (the $6^{th}$ to $8^{th}$, and $1^{st}$ phase frames) 
and two peaks appear during the last half period from the $2^{nd}$ to the $5^{th}$ phase frame (Figure 9 d).
 This is due to the amplitude of variation at the central 50 $\mu$as being twice as large as those of the surroundings.
 One peak appears when the central part becomes bright, and two peaks, one on the east and one on the west,
 appear when the central part darkens.
 Unlike other periods, the $P = 56.4$ min component seems concentrated and excited at the radius close to the black hole where the relativistic effect are not negligible, and hence it can be a key to distinguishing the origin of QPOs.

The detected pattern changes are not from noise effects: we simulated changes using artificial visibilities with noise levels showing the comparable amplitude fluctuations with those of real visibility data. 
 SMI images from the artificial visibilities show larger variations in not east-west but south-north direction that the spatial resolution is worse, while those from the observations show the opposite tendency.
 From the artificial visibilities, changes appeared in the image but none of the periodic regular patterns could be found. Also the artificial visibilities showed no coincident occurrence between spreading size minima and amplitude peaks in the spectra of the whole 209 area.
 Further, the observed periodic changes of real data basically did not disappear in SMI-maps with shorter total integration times (195 min) while different changes appeared from the artificial visibilities.

\section{DISCUSSION AND CONCLUSION}
 The radio QPOs reported here have a delay of 1.5 days from the millimeter wave flare, and the distributions extended to 100 \RS in apparent diameter while the QPOs in NIR and X~ray are just at the flaring time and estimated to occur around a marginally stable orbit (MSO)~(Genzel et al. 2003; Aschenbach et al. 2004; Eckart et al. 2006 and Belanger et al. 2006). The flaring events at MSO would vibrate the whole system to excite fundamental disk oscillation modes, which would last for several days at least.

There are features that cannot be explained as Keplerian motions directly governed by centripetal force, but that are easily interpretable as disk oscillation features~(Kato et al. 2008).
 First, the counter-rotations occurred simultaneously between the $P= 16.8$ min and $P = 22.2$ min periods, which is difficult to be explained by a single body motion like hot spot models.
 Second, the orbit of the shorter period $P = 16.8$ min is larger than that of the 
 $P = 22.2 min$, which cannot be realized by a Keplerian motion.

 If QPOs originate in a strong gravity field where the relativistic effect plays an important role, the periods of QPOs should depend on the mass and the spin of a massive black hole. Recent theories of disk seismology~(Kato et al. 2008; Abramowicz et al. 2001) predict that peak frequencies of QPOs can be scaled by a mass of central black holes as an analogy to QPOs in black hole x-ray binaries (BXB)~(Remillard, \& McClintock 2006). For example, in GRO J1655-40, a peak frequency of high frequency QPOs is about $3\times10^{2} (6.0 - 6.6 M_{\odot}/M_{\rm BH})$ Hz (where $M_{\odot}$ is a solar mass), with the result that a corresponding peak frequency using the mass of \SGR derived from the orbital motions of surrounding stars ($3.6\pm 0.3\times 10^{6}M_{\odot}$) ~(Eisenhauer et al. 2005)
is about $5.1\times 10^{-4} $ Hz ($P = 32 $min), which is one of our findings.
Detailed analysis with the obtained four QPO periods and wave-warp resonant oscillation model predicts the spin of \SGR to be $0.44\pm 0.08$ and the black hole mass to be $(4.2\pm 0.4)\times 10^{6}M_{\odot}$.~(Kato et al. submitted to MN)\\

 The apparent angular size of 0.1 mas corresponds to 10.4\RS(1\RS is Schwarzschild radius) assuming the Galactic center distance to be 7.6 kpc.~(Eisenhauer et al. 2005), which scale give us that the orbital velocities of bright peak positions are $v_{P=16.8 min}~=~2.1~c~at~8.8~Rs$ and $v_{P=22.2 min}~=~1.4~c~at~7.4~Rs$, both are superluminal, but can be explained as follows.
 It is well-known that the VLBI images of \SGR ~are obscured and broadened 
by the scatterings of surrounding plasma~(Lo et al. 1998; Doeleman et al. 2001).
 If we consider the broadening ratio of $2.6\sim3.0$ at 43 GHz by scattering~(Shen et al. 2005; Bower et al. 2004), and the magnifying ratio $\sim 1.23$ by self-gravitational lensing effect at 3 \RS, 
 the apparent 0.1 mas observed at 43 GHz is equal to 2.8 to 3.3 \RS in the intrinsic image of \SGR. Taking into these effects, we find that $v'_{P=16.8 min}~=~0.43~c~at~2.7~Rs$ and $v'_{P=22.2 min}~=~0.68~c~at~2.3~Rs$, which velocities are comparable to the Keplerian velocities at $r= 2-4 Rs$ from a black hole.\\

Presumably the emission region of our observed radio QPOs is different from those of other wavelength, even though the origin of oscillations
is common.  Kato et al. (2009) show that synthetic images of radio, NIR, and X-ray are quantitatively different, by using multi-wavelength
radiative transfer calculations including compton scattering process in three-dimensional magnetized hot accretion flows.  Without
synchrotron self-compton (SSC) process, Takahashi et al. (2009) show that temporal fluctuation of flux density at shorter wavelength,
where the disk becomes optically thin, is insensitive to the non-axisymmetric disk structure.  A possible emission mechanism of NIR
and X-ray wavelengths in order for having common oscillations at radio wavelength is the SSC, which is dominant when mass accretion rate
becomes large [see Fig. 4 (a) and (d) in Kato et al. 2009].\\

 The amplitude of the QPOs were 22 mJy ($P = 16.8$min), 24mJy ($P = 22.2$ min), 25mJy ($P = 31.4$min), and 20 mJy ($P = 56.4$ min) respectively, the total reached 90 mJy, nearly 10 \PC ~of the constant flux density component of the image (= 980 mJy). 
 The relative intensities of their periods are about $2~or~3 \PC$ level of the total flux density, which is comparable ratio to those of X~ray QPOs ($5 - 10 \PC$ level in case of X~ray Low Frequency QPO,
 $1\PC$ level in case of X~ray High Frequency QPO). 
  Such structural changes will cause fluctuations in fringe phase, rate and amplitude. Existence of oscillations with periods shorter than the whole observing time will be one of the reasons why the VLBI imaging of \SGR ~is difficult so far. 
 However at the same time, the detection of these periodic pattern changes means the obscured radio image of \SGR ~still bears the imprint of the intrinsic figure.
 The VLBA observations of SgrA$^{*}$, together with oscillation analysis, will provide us a unique chance to investigate the strong gravity field around the massive black hole SgrA$^{*}$ with order of \RS resolutions.

\section*{Acknowledgments}
 We would like to thank Shoji Kato, Ryoji Matsumoto, Shin Mineshige, Shinya Nitta,
 Mami Machida, Takahiro Kudoh, \& Hitoshi Negoro for discussions about disk oscillations, and Yoshiaki Tamura K. Y. Lo and James Moran for discussion about period analysis. The Very Long Baseline Array is operated by the National Radio Astronomy Observatory, which is a facility of the National Science Foundation, operated under cooperative agreement by Associated Universities Inc.

\appendix
\section{The SMI results around P = 12.9 min}

\begin{figure*}
\begin{center}
\includegraphics[height=0.55\textwidth]{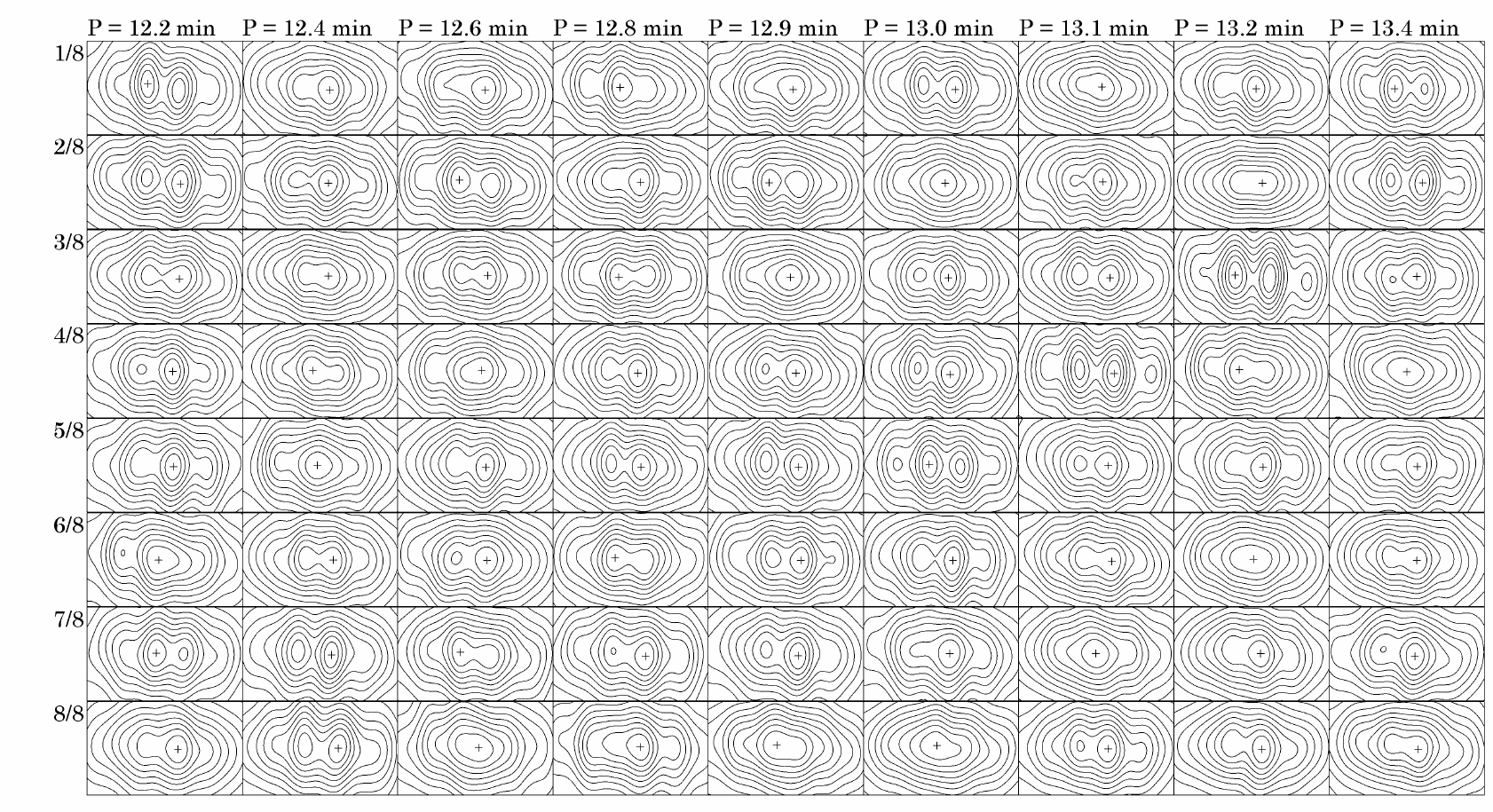}
\caption{ SMI search around P = 12.9 min.
 The slit-modulation-imaging~(SMI) maps with the 9 periods 
 $P = 12.2, 12.4, 12.6, 12.8, 12.9, 13.0, 13.1, 13.2, and 13.4 $min.
 The 8 maps show the images in phase frame series with respective periods.
 The cross mark shows the position of peak intensity in each map.
 The contour levels are every $10 \PC $~of the peak brightness of the respective maps.
 From these trial periods we detected no conspicuous pattern change with the period.}
\label{fig SMI129}
\end{center}
\end{figure*}

We here show the none detection cases of conspicuous pattern change with the trial periods using the SMI method. As shown in Figure 8 (f), (g), there is a small peak at the period P = 12.9 min in the power spectra.
Also the Figure 8 (h) shows that the spreading size of the P = 12.9 min is relatively small.
Using the SMI method we searched around the period for any conspicuous pattern changes.
One-peak or two-peaks structures appeared in the phase frames of SMI maps of the respective trial periods.
Unlike the cases of the P = 16.8, 22.2, 31.4, and 56.4 min, the change patterns in the phase frames seem at random.
We therefore could not confirm the QPO detection around P = 12.9 min.

\section{Spectra from smaller array}
 The QPO spectra can be detected from images with limited \UVC ~coverage.
 In order to indicate that the obtained QPO spectra are not due to insufficient calibrations
of the stations of MK, SC, and HN,
 we produced series of 5 minutes integration snap shot maps without the visibilities of the longer baselines,
 and performed the same analysis of the time variations of flux densities in
 several regions of the mapping area.\\
 We omitted the following baselines from the original calibrated data. \\
\begin{itemize}
\item All the baselines to the MK station.
\item The 7 baselines to the SC stations, namely the baselines of SC-BR, SC-FD, SC-KP,SC-LA,
 SC-NL, SC-OV, and SC-PT. We keep the data of the SC-HN baseline because the visibilities
 of SC-HN show quite high SNR.
\item The 3 baselines to the NL station, namely NL-BR, NL-OV, and NL-SC baselines.
\end{itemize}
It is already noted in the section 2 that the original calibrated data
 do not contain the baselines of MK-SC, and MK-HN. 
 Figure B1 (b) shows the \UVC ~coverage of the limited data set.
 For comparison, we show the whole \UVC ~coverage of the original calibrated data in Figure B1 (a),
 which is the same figure as Figure 1. \\
  Figure B2 (a) is the same image as shown in Figure 8 (a) with a different contour expression,
while Figure B2 (b) shows the image of the limited data set from the whole time integration.
We here used the common restoring beam of $0.40 mas \times~0.15 mas, PA = 0 \DEG$.
Other parameters of the imaging are the same as written in the section 2-5.
Because of the poor \UVC ~coverage of the limited data,
 the resultant image is a simple obuje only representing the size and 
the elongation direction (east-west) of the \SGR shape at 43GHz.

From the series of snap shot maps, we measured the flux densities in the 8 regions defined
as shown in Figure B3, and analyzed the time variations using MEM and found the power spectra of periodicity
shown in Figure B4. There are two facts to be noted.
First, both of the spectra from the whole \UVC ~data (black lines) and the limited \UVC ~data (red lines)
show quite similar curve to each other in respective regions.
Second, peaks appear around P = 17, 22, 30, 55 min in the spectra of the limited \UVC ~data.
 These two facts indicate at least that the detected QPOs are not due to insufficient calibrations
of the stations of MK, SC, and HN.

\begin{figure*}
\begin{center}
\includegraphics[height=0.4\textwidth]{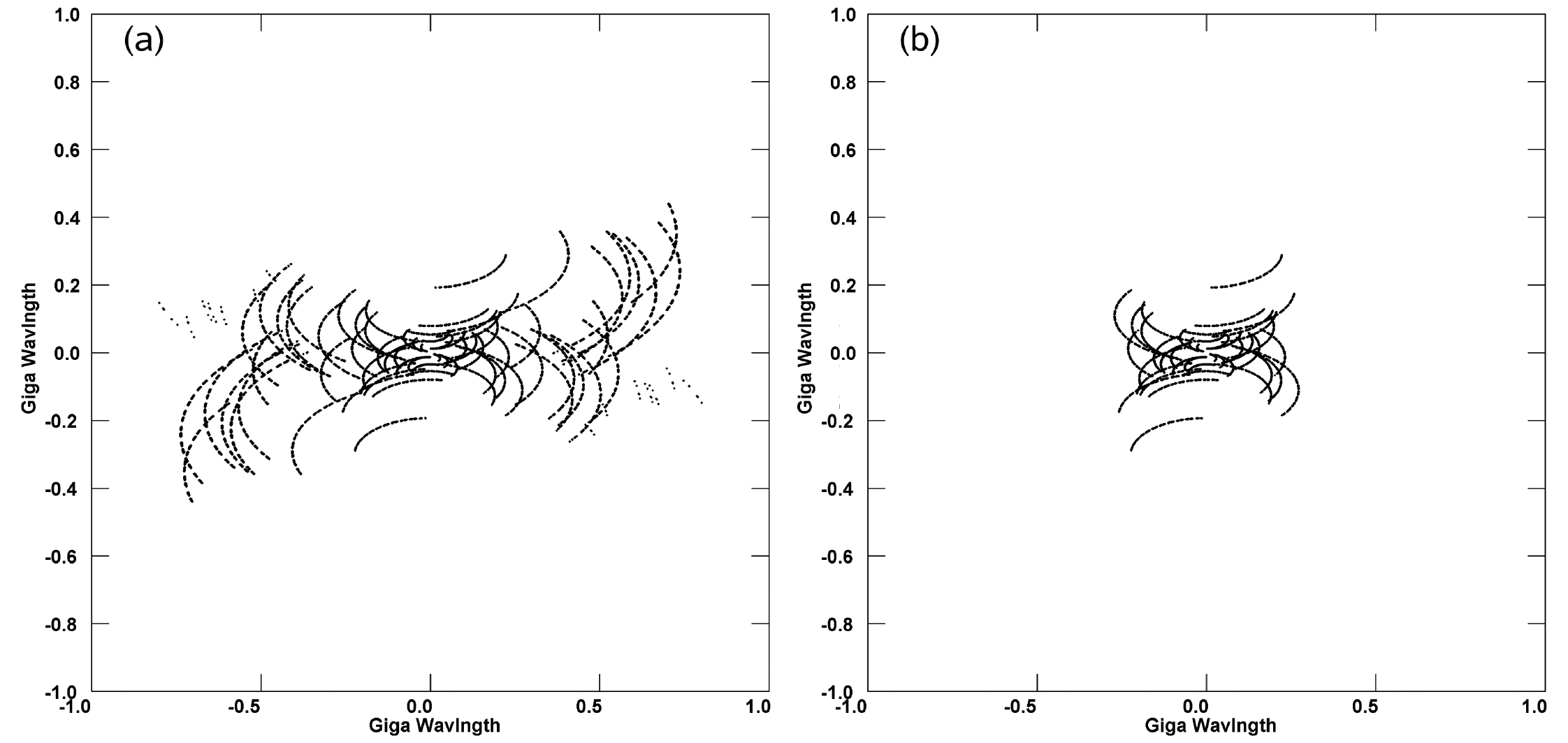}
\caption{The \UVC ~coverages of the whole calibrated data (a) and the limited data (b).
The Figure B 2 (a) is the same plot as Figure 1. }
\label{figsmalluv1}
\end{center}
\end{figure*}
\begin{figure*}
\begin{center}
\includegraphics[height=0.5\textwidth]{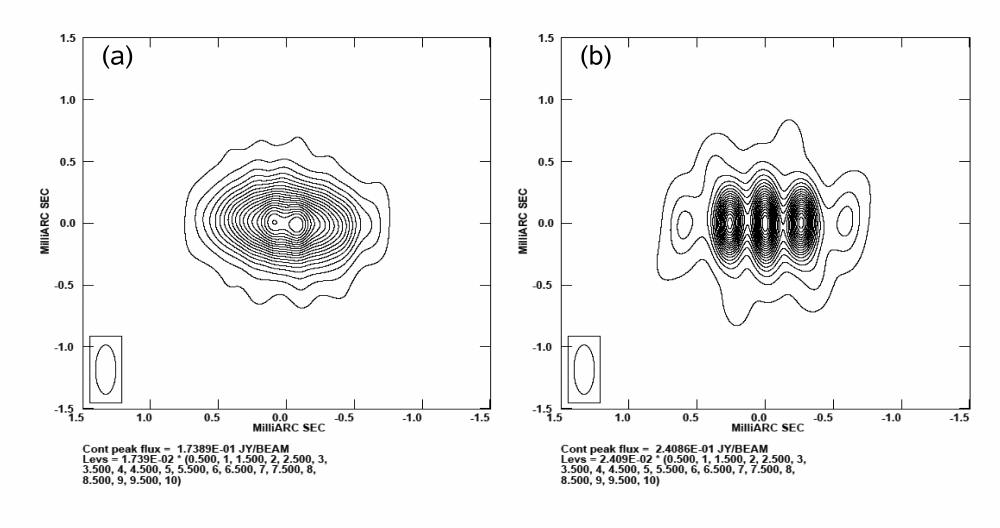}
\caption{maps}
\label{The resultant images from the whole calibrated data (a) and the limited data (b).
The Figure B 3 (a) is the same image as shown in Figure 8 (a), but with different contour expression.}
\end{center}
\end{figure*}
\begin{figure*}
\begin{center}
\includegraphics[height=0.55\textwidth]{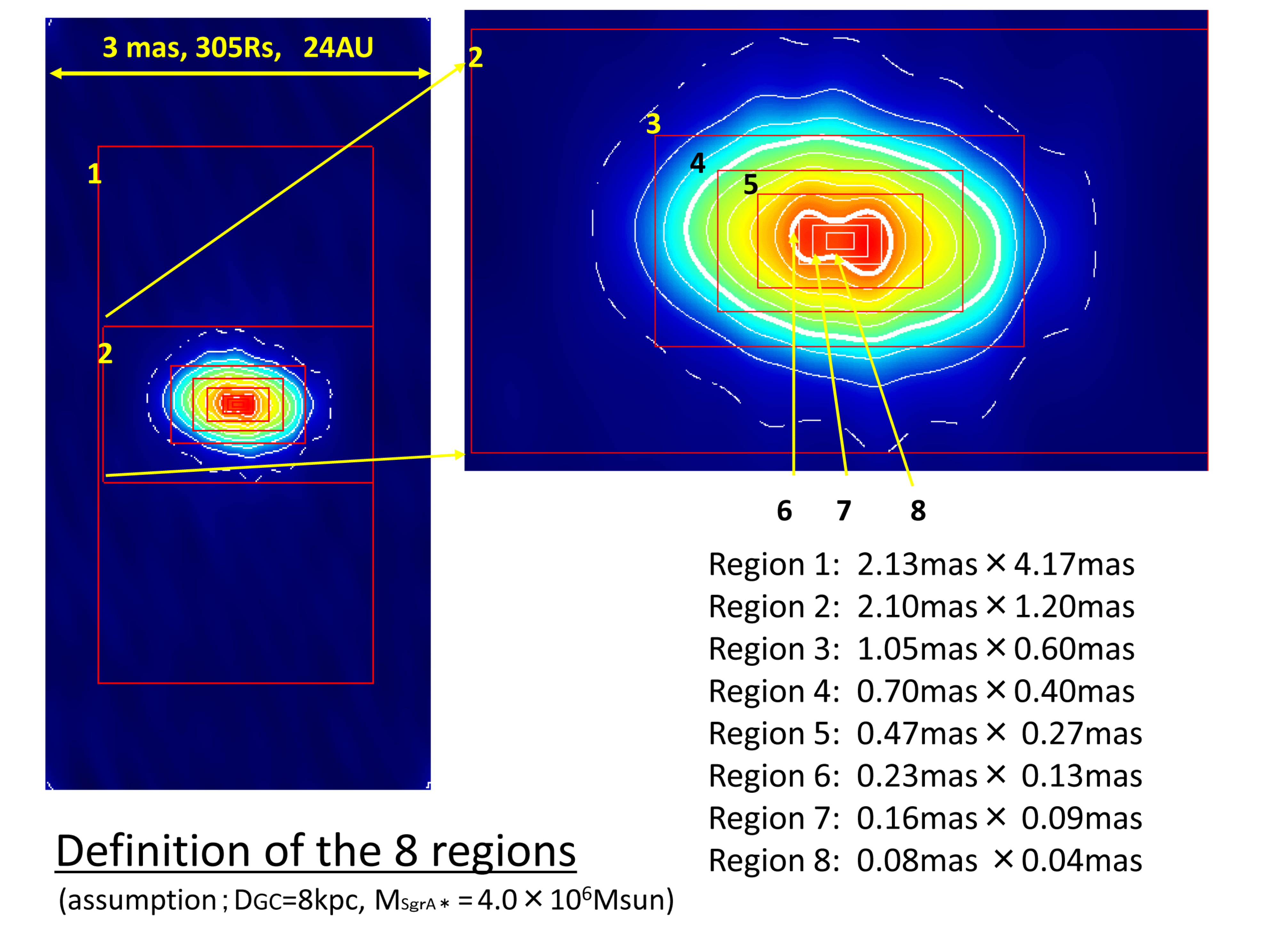}
\caption{The definition of the 8 regions. The size of region (4) is comparable to so called a
scattering size of \SGR at 43GHz. Assuming the distance to the Galactic center to be 8 kpc and the mass
of \SGR to be $4 \times 10 ^{6} M_{sun}$, the apparent 3 mas corresponds to be 24 AU, or
305 Schwarzschild radii.}
\label{8regions}
\end{center}
\end{figure*}
\begin{figure*}
\begin{center}
\includegraphics[height=0.55\textwidth]{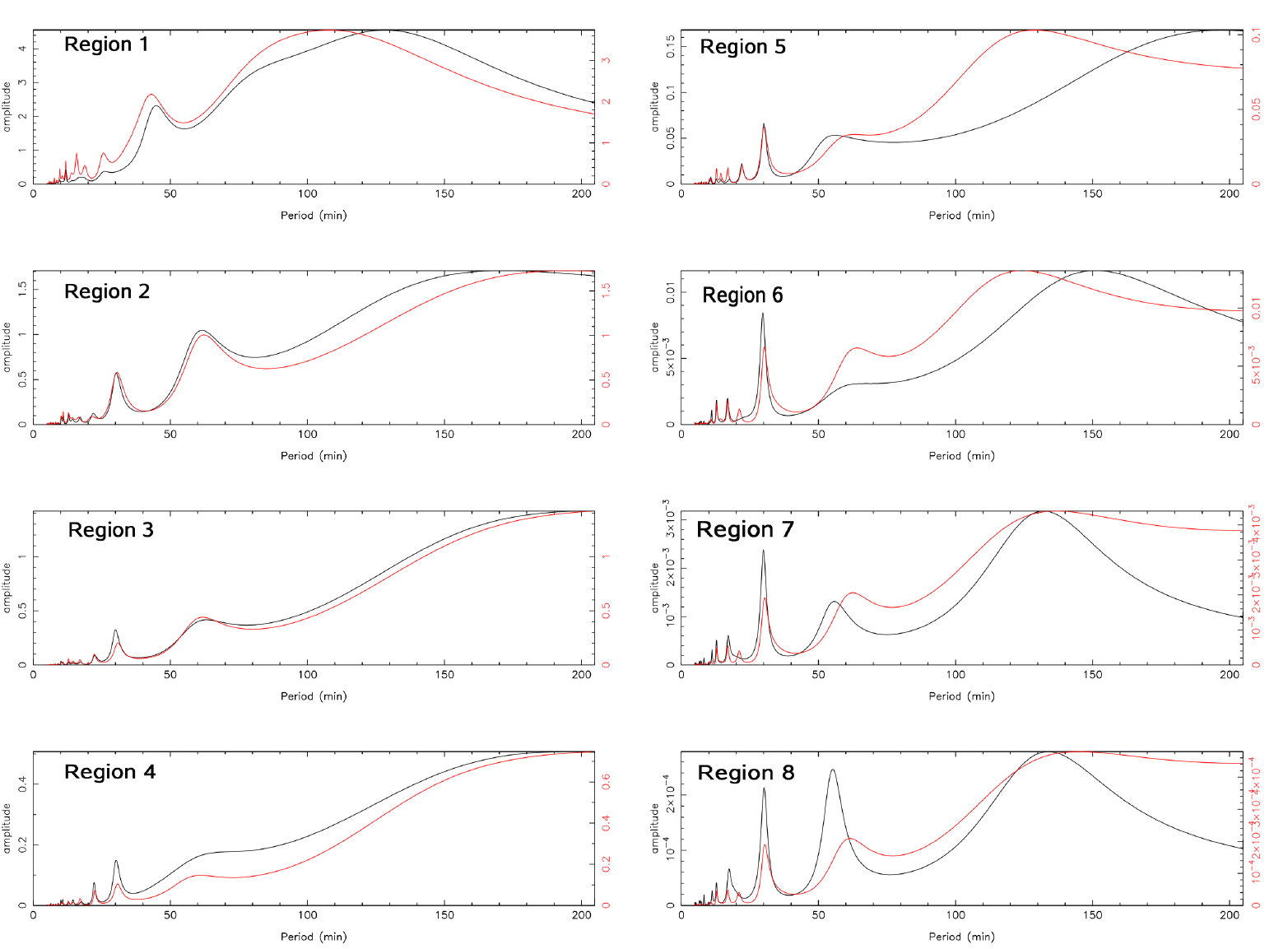}
\caption{The power spectra of the periodicities in the 8 regions.
Black line shows the spectrum obtained from the whole calibrated data, red line
shows the spectrum obtained from the limited calibrated data.}
\label{spactracomparison}
\end{center}
\end{figure*}
\section{Calculated closure phases from the SMI maps (P =16.8 min)}
In Figure C 1 and 2, we show the closure phases calculated from the 8 maps from the SMI method at P= 16.8 min. The SMI maps show consistent closure phases to the observed ones.
First of all, the small triangles containing FD, LA, PT, KP, and OV stations shows nearly zero, constant
closure phases both in observed closures and the calculated ones.
Second, the observed closure phases of triangles containing one station from BR or NL and two stations 
from the inner five stations FD, LA, PT, KP, and OV show deviations of a few tens of degrees from zero.
 The calculated closure phases from the SMI maps show different values from the observed ones, however,
 the distributions of them cover the regions of variations of the observed closure phases.
Third, the larger the triangles become, the observed closure phases show larger deviations up to
 $\pm 180 \DEG$. The calculated closure phases from the SMI maps also show 
the same large deviations.
From the point of closure phases, the SMI maps show consistent behavior to the observed closure phases.

\begin{figure*}
\begin{center}
\includegraphics[height=1.40\textwidth]{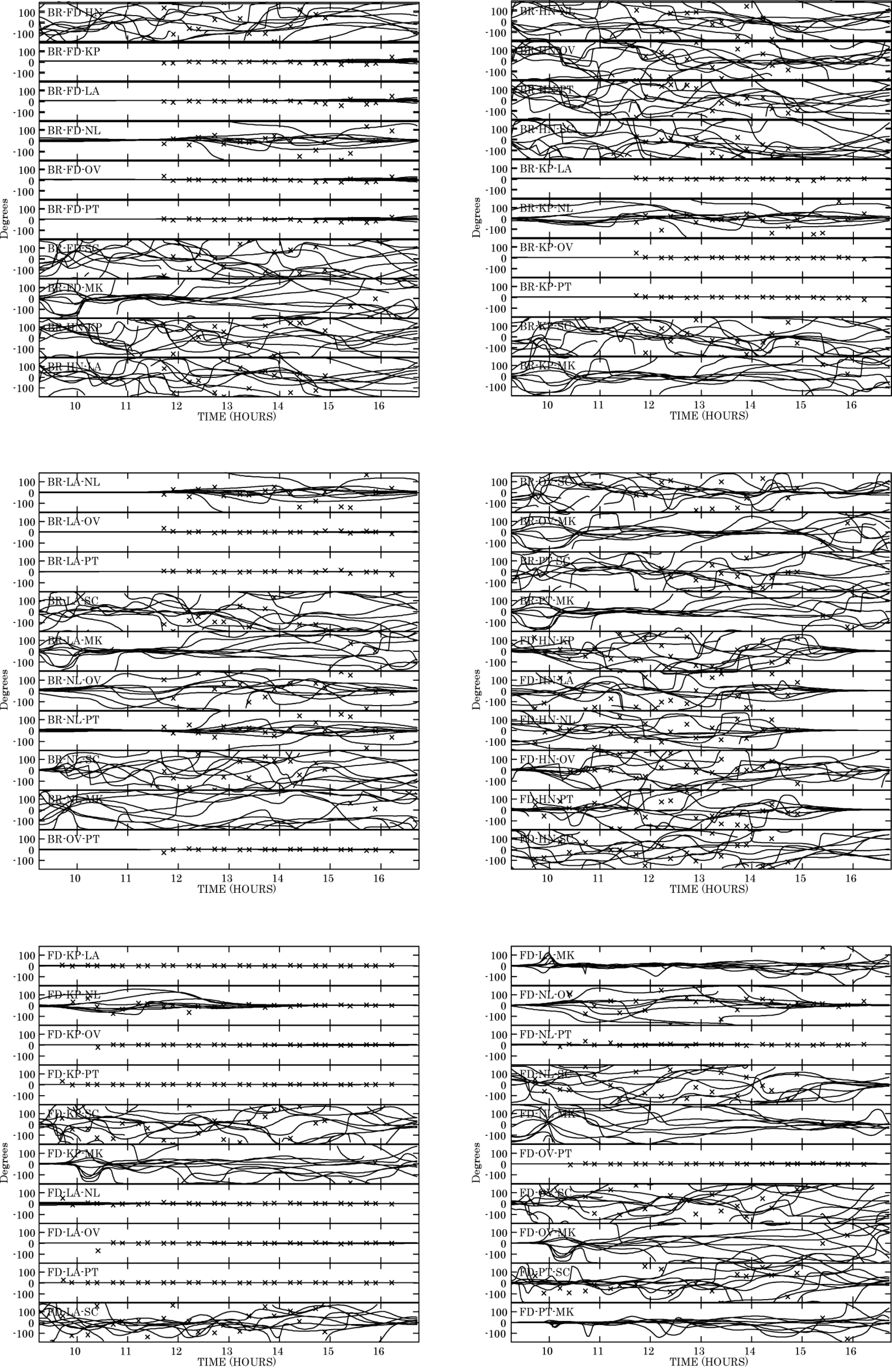}
\caption{The calculated closure phases from the eight SMI maps (P =16.8 min) (1/2).}
\label{closurephaseSMI1}
\end{center}
\end{figure*}
\begin{figure*}
\begin{center}
\includegraphics[height=1.20\textwidth]{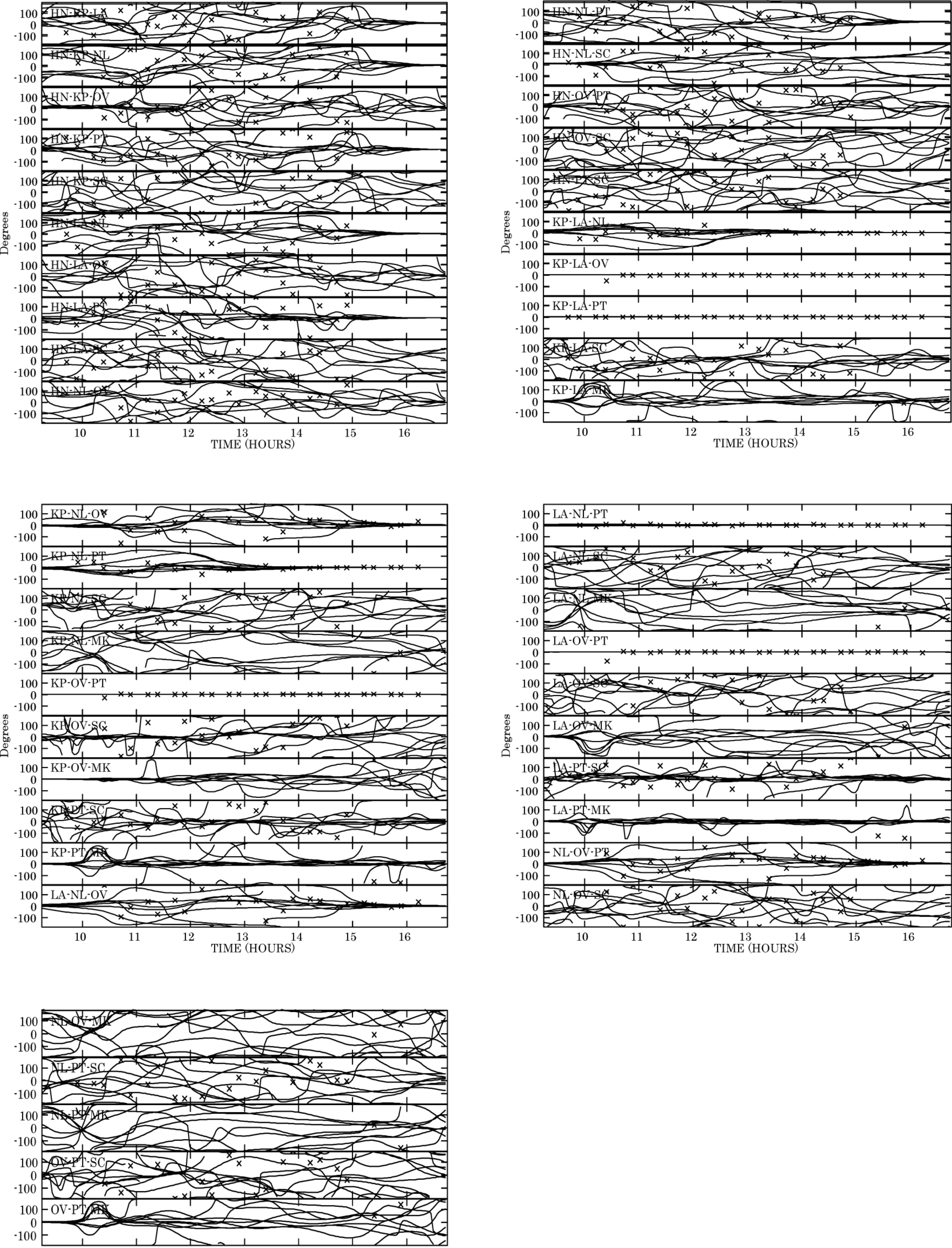}
\caption{The calculated closure phases from the eight SMI maps (P =16.8 min) (2/2)}.
\label{closurephaseSMI2}
\end{center}
\end{figure*}
\section{Images with several limited visibility sets}
 Figure D shows several images from the whole time integration with several limited visibility sets.
Here we performed imagings with the task IMAGR with the same parameters as described in the section
2-5.
 Visibility data from the baselines to MK seem to do almost nothing to the image quality.
The 4 images without the baselines to MK, (b), (d), and (e) in Figure D, are very similar to each other
and also to the image from all data, (a) in Figure D.
 While the existences of the baselines to HN, and SC in data play fairly an important role to the imaging quality. The exception of one from the two stations gives little influence to the image quality ((c), (d), (e), (f), (g) in Figure D). The exception of both of the HN, SC stations gives quite a degree of influence in image quality ((g), and (h) in Figure D). 

\begin{figure*}
\begin{center}
\includegraphics[height=0.50\textwidth]{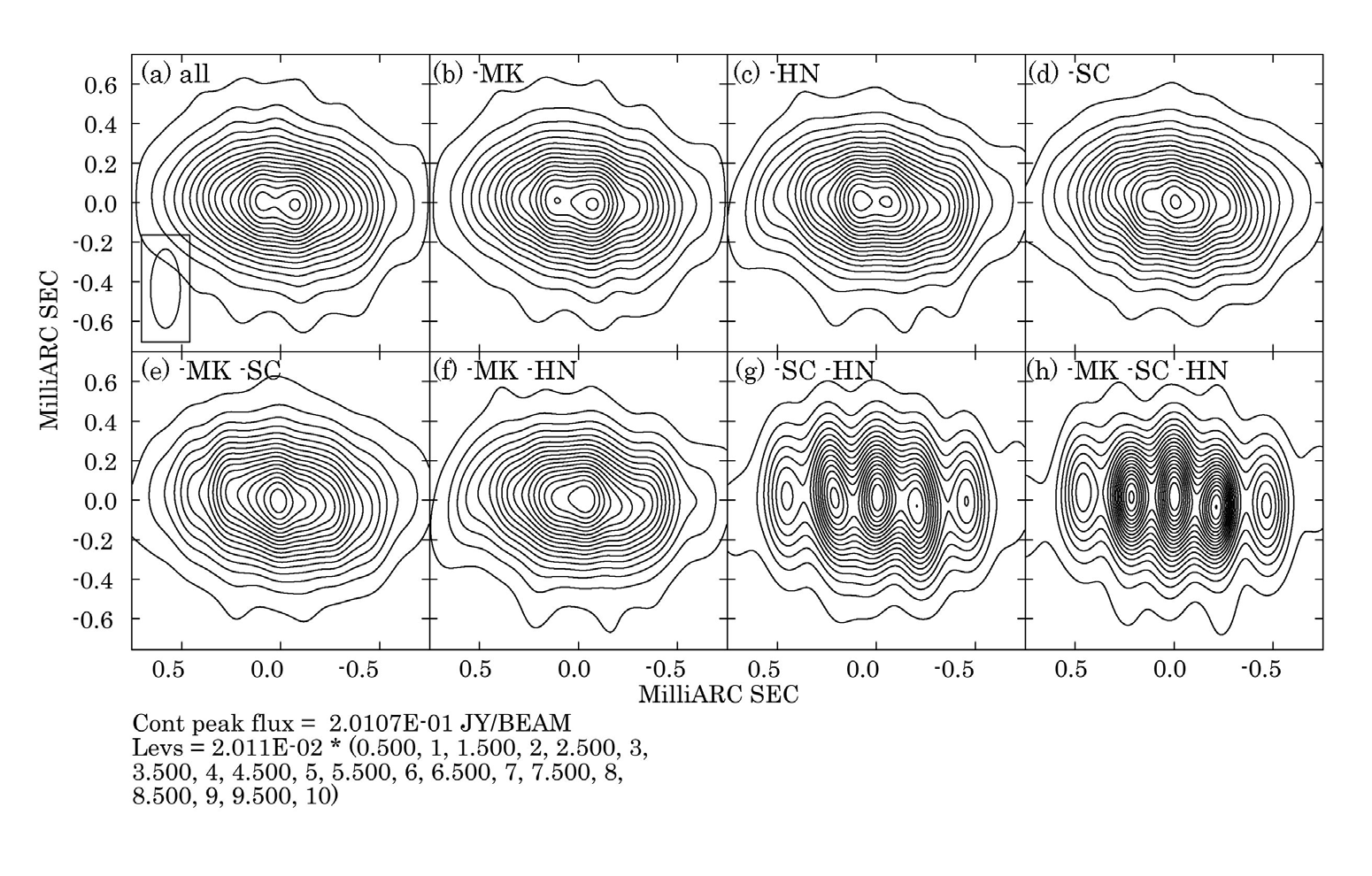}
\caption{Several images with limited visibility data.
(a) the image using all the calibrated data, the same one as shown in Figure 8 (a).
(b) an image using all the calibrated data except baselines to MK station.
(c) an image using all the calibrated data except baselines to HN station.
(d) an image using all the calibrated data except baselines to SC station.
(e) an image using all the calibrated data except baselines to MK and SC stations.
(f) an image using all the calibrated data except baselines to MK and HN stations.
(g) an image using all the calibrated data except baselines to SC and HN stations.
(h) an image using all the calibrated data except baselines to MK, SC, and HN stations.
}
\label{maps661}
\end{center}
\end{figure*}

\end{document}